\documentclass[12pt]{iopart}
\pdfoutput=1
\usepackage{color}
\usepackage[applemac]{inputenc}
\usepackage{amssymb}
\usepackage[export]{adjustbox}
\usepackage{comment}
\usepackage{amsfonts}
\usepackage{amssymb}
\usepackage[autostyle]{csquotes}
\usepackage{graphicx}

\def\beqra{\begin{eqnarray}}
\def\eeqra{\end{eqnarray}}
\def\beq{\begin{equation}}
\def\eeq{\end{equation}}

\def\qm{q_{\rm{max}}}

\def\bx{{\bf{x}}}

\def\bk{{\bf{k}}}
\def\bp{{\bf{p}}}
\def\bq{{\bf{q}}}
\def\bd{{\bf{d}}}

\def\bv{{\bf{v}}}

\def\bV0{{\bf{V_0}}}
\def\re#1{(\ref{#1})}

\def\alt{~\mbox{\raisebox{-.6ex}{$\stackrel{<}{\sim}$}}~}
\def\bx{{\bf{x}}}

\def\br{{\bf{r}}}

\def\bk{{\bf{k}}}
\def\bp{{\bf{p}}}
\def\bq{{\bf{q}}}

\def\bv{{\bf{v}}}

\def\hmp{\rm{h\,Mpc^{-1}}}
\def\bz{{\bf{z}}}

\begin{document}

\begin{flushright}
        \quad \\
        \quad \\
        \quad \\
        YITP-20-49\\
\end{flushright}

\title[Model independent measurement of the growth rate from the CR's of the LSS]{Model independent measurement of the growth rate from the consistency relations of the LSS
}

\author{Marco Marinucci}
\vskip 0.3 cm
\address{
Dipartimento di Scienze Matematiche, Fisiche ed Informatiche dell'Universit\`a di Parma, Italy\\
and INFN, Sezione di Parma\\
}
\author{Takahiro Nishimichi}
\address{
Center for Gravitational Physics, Yukawa Institute for Theoretical Physics, Kyoto University, Kyoto 606-8502, Japan
}
\address{
Kavli Institute for the Physics and Mathematics of the Universe (WPI), UTIAS, The University of Tokyo, Kashiwa, Chiba 277-8583, Japan
}
\author{Massimo Pietroni}
\address{
Dipartimento di Scienze Matematiche, Fisiche ed Informatiche dell'Universit\`a di Parma, Italy
}

\begin{abstract}
The Consistency Relations for the Large Scale Structure provide a link between the amplitude  of Baryonic Acoustic Oscillations in the squeezed bispectrum (BS) and in the power spectrum (PS). This relation depends on the large scale bias of the considered tracer,  $b_\alpha$, and on the growth rate of structures, $f$. Remarkably, originating from basic symmetry principles, this relation is exact and independent on the underlying cosmological model.

By analysing data from large volume simulations, both for dark matter and for haloes, we illustrate how BS and PS measurements can be used to extract  $b_\alpha$ and $f$  without the need of any theoretical approximation scheme  for the computation of the BS and the PS. We show that, combining measurements of the squeezed BS with the quadrupole to monopole ratios for the PS at large scales can successfully break the $b_\alpha -f$ degeneracy. We forecast that this method, applied to a Euclid-like survey, will be able to measure bias, and then the growth rate, at better than $10\%$ level, with no extra assumption.

\end{abstract}

\maketitle

\section{Introduction}
Next generation Large Scale Structure (LSS) surveys will measure the parameters of the present standard model of cosmology, namely $\Lambda$CDM, to an unprecedented precision. On the theoretical side, this requires a) setting up a computational prescription to compute the relevant observables, based on numerical simulations and/or semi-analytical approaches, and, b) exploring the parameter space of the model to find the most probable regions given the data. Consistency Relations for the LSS (CR's) \cite{Peloso:2013zw, Kehagias:2013yd} provide a way to extract cosmological information in a way independent both on a computational scheme, and on the cosmological model, and therefore they represent a potentially unique way to perform unbiased tests. In ref.~\cite{Marinucci:2019wdb} we showed it explicitly in real space, demonstrating the potential of CR's in measuring the linear bias parameter, which relates the distributions of a given tracer to that of Dark Matter (DM) at very large scales. In this paper we take a step forward, exploring the CR's in redshift space as a way to measure the growth function at different redshifts in a model independent way, a crucial cosmological test.

Physically, the CR's account for the contribution to the bispectrum (BS) induced by  matter displacements coherent on very large scales. As such, this contribution can be disentangled from the other ones -- induced by the various sources of nonlinearities at play -- by looking at the squeezed BS limit, in which the modulus of one of the three wavevectors, $q$, is much smaller than the other two, of order $k \gg q$. The properties of the large scale displacements are dictated by the relevant symmetry of the system, that is the Equivalence Principle (EP), and by the properties of the initial conditions. Assuming adiabatic and gaussian initial conditions, CR's single out  a nonperturbative contribution to the squeezed limit BS, where by `nonperturbative' here it is meant that it does not rely on any approximation scheme (like, for instance Perturbation Theory (PT)) and that the result holds even beyond the perfect fluid approximation. By further assuming that the very `long' modes at scale $q$ can be described by linear PT (but with no approximation on the `short' ones at scale $k$!) the CR's take the form of exact relations between the BS and the PS's evaluated at the scales $k$ and $q$ (see Eq.~\re{RSder} below). For the above reasons, the CR's hold not only for matter but for any tracer, both in real and in redshift space \cite{Peloso:2013spa, Creminelli:2013poa}.
The potential of CR's in constraining possible violations of gaussianity of the initial conditions has been investigated in  \cite{Valageas:2016hhr, Esposito:2019jkb}, while for violations of the EP see \cite{Peloso:2013spa,Creminelli:2013nua}.

At first sight, for theories respecting the EP and assuming gaussian initial conditions, the nontrivial content of the CR's might look empty, as in this case the contribution to the BS `protected' by the CR's is parametrically of the same order, $O((k/q)^0)$, of other terms induced by different sources of nonlinearities at any PT order, for which theoretical control is limited by the reliability of PT in the considered range. However, as it was shown in \cite{Marinucci:2019wdb}, the wiggly feature of Baryonic Acoustic Oscillations imprinted in the spectra of the tracers of LSS in the late-time universe provide a way to isolate the CR-protected contribution to the PS from unprotected ones. The latter, although being parametrically of the same order in $k/q$ in the squeezed limit, are either smooth or suppressed by factors  $O(2 \pi / (k r_s))$ with respect to the protected ones, where $r_s
\simeq (100 \, {\rm Mpc\,h^{-1}})$
is the BAO acoustic scale. Therefore, in real space, by comparing the BAO amplitudes in the BS and in the PS we were able to measure the prefactor of the latter which is related to the linear bias parameter. After fifteen years of the first clear detection of BAO in the galaxy correlation function by \cite{Eis05}, the precision at which the BAO feature can be extracted from the distribution of galaxies has greatly been improved (see \cite{Alam:2016hwk} for recent results from the PS and the correlation function). Furthermore, a high-significance detection of the BAO feature in the three-point functions has also been reported by \cite{Slepian_2017}. Therefore, we expect that measurements of the CR's through the BAO feature in real data would be within the reach of large-scale experiments planned in the near future, such as LSST \cite{Tyson:2002nh}, Euclid \cite{Amendola:2016saw}, WFIRST \cite{Dore:2019pld}.

In this paper we extend the analysis of \cite{Marinucci:2019wdb} to redshift space, in which the CR's coefficients depend both on the large scale bias and the large scale  growth function, $f= d\ln D/d\ln a$ (where $D$ is the linear growth rate) thereby providing a  way to break the degeneracy between the two. We will analyse large volume N-body simulations, confirming the validity of CR's also in redshift space. We will find  that CR's alone are mostly sensitive to the bias $b_\alpha$  (where $b_\alpha$ is the linear bias of  the tracer $\alpha$ ), giving weak constraints on the parameter $\beta_\alpha=f/b_\alpha$. However, by  combining the CR analysis with the independent extraction of the parameter $\beta_\alpha$ from the PS quadrupole to monopole ratio, the
$f-\beta_\alpha$
degeneracy can be successfully broken.  Moreover, we estimate the constraining potential of future surveys, in particular, Euclid \cite{Amendola:2016saw}, showing that it can reach better than $10\%$ precision on the bias parameter, and therefore on $f$ as well, in a manner completely free from assumptions on the biasing prescription as well as the underlying gravity theory.

The paper is organised as follows. In Sect.~\ref{CRdisc} we obtain the CR's in redshift space for biased tracers; in Sect.~\ref{mp} we define the multipoles of the BS and PS's and obtain CR's in terms of these; in Sect.~\ref{sims} we describe the set of simulations we  use for the analysis of this paper and the procedure  used to measure the BS and the PS both for DM and for halos of different masses at different redshifts; in Sect.~\ref{resid} we describe our analysis and present its results on the bias parameters and the growth function; in Sect.~\ref{forecast} we estimate the costraining power of this methodology when applied to future data from the Euclid survey. Finally in Sect.~\ref{concl} we summarize our conclusions and give our outlook on future developments. In \ref{gender} we give details on the derivation of the CR's for biased tracers in redshift space.

\section{Consistency relations: biased tracers in redshift space}
\label{CRdisc}
The equal-times CR for a single tracer, $\alpha$,  in redshift space (for a general derivation, see \ref{gender}), reads,

\beqra
 &&\!\!\!\!\!\!\!\!\!  \!\!\!\!\!\!\!\!\!  \!\!\!\!\!\!\!\!\!\!\!\!\!\!\! \lim_{q/k\to0} \frac{B_{\alpha}(\bq,-\bk_+,\bk_-)}{P_{\alpha}(\bq) P_\alpha(\bk)} \nonumber\\
 &&\!\!\!\!\!\!\!\!\!\!\!\!\!\!\!\!\!\!\!\!\!\!\!\!\!\!\!\!\!\!\!\!= \lim_{q/k\to0}-\frac{k}{q}\frac{\mu + f \mu_k \mu_q}{b_\alpha + f\mu_q^2}  \frac{P_\alpha(\bk_+) - P_\alpha(\bk_-)}{P_\alpha(\bk)}+ O\left(\left(\frac{q}{k}\right)^0\right)\,,
 \nonumber\\
 &&\!\!\!\!\!\!\!\!\!\!\!\!\!\!\!\!\!\!\!\!\!\!\!\!\!\! \!\!\!\!\!\! = - \frac{\mu^2 + f\mu\mu_k\mu_q}{b_\alpha + f\mu_q^2}\frac{\partial \ln P_\alpha(\bk)}{\partial \ln{k}} - \frac{\mu_k(\mu + f\mu_k\mu_q)(\mu_q - \mu \mu_k)}{b_\alpha + f\mu_q^2} \frac{\partial \ln P_\alpha(\bk)}{\partial \ln{\mu_k}}
+ O\left(\left(\frac{q}{k}\right)^0\right)\,,
 \label{RSder}
\eeqra
where $k\equiv |\bk|$, $q\equiv |\bq|$, $\bk_{\pm}\equiv \bk\pm \bq/2$, $\mu \equiv \bk\cdot\bq/(k\,q)$,  $\mu_k\equiv \bk\cdot \hat z/k$, and  $\mu_q\equiv \bq\cdot \hat z/k$, with $\hat{z}$ being the direction of the line of sight. We have omitted the time dependence and used the fact that, in the far observer approximation, the redshift space PS, $P_\alpha(\bk)$, depends only on $k$ and $\mu_k$. In the following, we will consider dark matter ($\alpha=m$), halos ($\alpha=h$), and galaxy ($\alpha=g$) tracers.

 We stress that the `linear bias' $b_\alpha$ appearing in the CR is not a parameter of a bias expansion, but is defined precisely as the limit between the real space PS for the tracer $\alpha$ and the $\alpha$-matter cross-correlator (see \cite{Marinucci:2019wdb} and \ref{gender}),
\beq
b_\alpha \equiv \lim_{q\to 0} \frac{P_{\alpha \alpha}(q)}{P_{\alpha m}(q)}\,,
\label{bias}
\eeq
the only assumptions entering this definition being the EP and adiabatic initial conditions, who ensure that all species move with the same velocity fields at large scales, and that linear PT holds at scales $q\to 0$.

Unlike the non-equal times CR's \cite{Peloso:2013spa}, the CR-protected contributions on the RHS of the equal times CR, namely the first term on the first line and the first two terms at the second one, cannot be distinguished from the unprotected ones (the $O((q/k)^0)$ terms), by looking at a pole in $q$ as the squeezed limit is approached. This is, at first sight, unfortunate, as the equal-times BS is, differently from the unequal-times one, not measurable from data. However, as it was discussed in \cite{Marinucci:2019wdb} and will be elaborated on in the following, BAO oscillations provide a way to single out the CR protected terms.

Among the two terms at the last line of Eq.~\re{RSder}, there is a hierarchy. The amplitudes of the oscillations in the logarithmic derivative of the PS with respect to $k$ are enhanced with respect to the ones of the derivative with respect to $\mu_k$.
 This can be understood by looking at models for the redshift space PS, such as \cite{Taruya:2010mx}, which can be cast in the form
\beq
P_\alpha(\bk) \simeq F_{fog} (k \mu_k f \sigma_v) (b_\alpha + f\mu_k^2)^2  \left(P^0(k)  + \Delta P_\alpha^{\rm{1-loop}}(k,\mu_k)\right)\,,
\label{fog}
\eeq
where $F_{fog}(x)$ is a phenomenological smooth function, usually a gaussian or a lorentzian, $P^0(k)$ is the linear PS, and  $ \Delta P_\alpha^{\rm{1-loop}}(k,\mu_k)$ are contributions of 1-loop order.
If we write
\beq
P^0(k)=P^0_{\rm{nw}}(k)\left(1+A(k) \sin(k r_s)\right)\,,
\eeq
where $P^0_{\rm{nw}}(x)$ is the smooth component of the linear PS, and $r_s=O(100)\,\hmp$ is the BAO scale, we see that the oscillating part of the logarithmic derivative of the PS with respect to $\ln k$ is of order
\beq
\frac{\partial \ln P_\alpha(\bk)}{\partial \ln k} \sim A(k)\, kr_s\,\cos(k r_s)+\rm{smooth/higher\;orders\;\;contributions}\,,
\label{lead}
\eeq
while
\beq
\!\!\!\!\!\!\!\!\!\! \!\!\!\!\!\!  \frac{\partial \ln P_\alpha(\bk)}{\partial \ln \mu_k} \sim\frac{\partial \Delta A(\bk)}{\partial \ln \mu_k} \sin(k r_s)+\rm{smooth/higher\;orders\;\;contributions}\,,
\eeq
where $\Delta A(\bk)$ is the   $\mu_k$-dependent part of the 1-loop contribution to the amplitude of the oscillating part of the PS. Therefore, comparing the two oscillating contributions, we see that besides being of 1-loop order as opposed to linear, the latter is suppressed by an extra factor of order $1/k r_s$,
\beq
\frac{1}{k r_s}\frac{\Delta A(\bk)}{A(k)} = \frac{k_s}{2 \pi k}\frac{\Delta A(\bk)}{A(k)} \,,
\label{sup}
\eeq
where we have defined $k_s=2 \pi r_s^{-1} \simeq 0.06\, \hmp$, and can then be safely neglected in the BAO range of scales for squeezed configurations.

Coming now to the oscillating part of the $ O\left(\left(\frac{q}{k}\right)^0\right) $ terms in Eq.~\re{RSder}, a perturbative analysis shows that they are of order
\beq
\frac{\Delta P_\alpha(k)}{P^0_{\rm{nw}}(k)} A(k) \sin(k r_s)\,,
\eeq
where $\Delta P_\alpha(k)$ is a one-loop order contribution to the PS. Therefore, compared to the leading oscillatory contribution, \re{lead}, this one is parametrically 
suppressed as the ones in \re{sup}, and therefore will be neglected too. In Sect.~\ref{resid} we will verify, from simulations,  that the difference between the LHS and the first term at the RHS of \re{RSder} is indeed smooth in the squeezed limit.

Summarising, in our analysis we will consider only the first term of Eq. \re{RSder}, that is,
\beqra
&&\lim_{q/k\to 0}B_\alpha(\bq,\bk_-,-\bk_+) = - \frac{\mu^2 + f \mu \mu_k \mu_q}{b_\alpha + f\mu_q^2} P_\alpha(\bq)\frac{\partial P_\alpha(\bk)}{\partial\ln{k}} \nonumber\\
&&\qquad\qquad\qquad\qquad\quad+\rm{smooth/higher\;orders\;\;contributions}.
\label{CRRS}
\eeqra

\section{Multipoles}
\label{mp}

We will deal with the angular dependence of the BS by considering multipole expansions. In the following, we discuss the redshift space case, Eq.~\re{CRRS}, from which the real space results can be derived by taking $f\to 0$ and reinterpreting the PS's and the BS's as being the real space ones.

The BS in redshift
space (in the distant observer approximation) depends on 5 coordinates: 3 of them (for instance $q$, $k$, and $\mu$) identify the triangular shape, while the remaining 2 are needed to define the orientation of the plane of the triangle with respect to the line of sight. Therefore, keeping $q$ and $k$ fixed, we are left with $3$ angular coordinates,
over which we will integrate with the measure
\beq
\int\frac{d^2\hat k}{4 \pi}\int\frac{d^2\hat q}{4 \pi}\,(2\pi) \delta_D\left(\phi_k+\phi_q \right) = \int_{-1}^1\frac{d \mu_k}{2}\int_{-1}^1\frac{d \mu_q}{2} \int_0^{2\pi}\frac{d\phi}{2\pi}\,,
\label{measure}
\eeq
where the delta-function in the first integral enforces rotation invariance around the $z$-axis, and we have defined $\phi=\phi_k-\phi_q$. The cosine $\mu$ is given in terms of the three independent variables as
\beq
\mu=\mu(\mu_k,\mu_q,\phi)=\sqrt{(1-\mu_k^2)(1-\mu_q^2)} \cos \phi+\mu_k\mu_q.
\label{murel}
\eeq

The PS's will be expanded in Legendre polynomials, ${\cal P}_l(\mu_k)$, as usual,
\beq
P_\alpha(\bk)=\sum_{l=0}^\infty \,P_\alpha^{(l)}(k)\,{\cal P}_l(\mu_k)\,,
\eeq
where
\beq
P_\alpha^{(l)}(k) \equiv \frac{2 l+1}{2} \int_{-1}^1 d\mu_k \, P_\alpha(\bk) \,{\cal P}_l(\mu_k)\,.
\eeq
Concerning the PS at the large scale $q$, $P_\alpha(\bq)$, we will use the same expansion as above, with the additional assumption that linear PT holds at the scale $q$, which is consistent to what we have already assumed in deriving the CR. This implies that the Kaiser relation   \cite{Kaiser:1987qv} can be used for the PS at this scale,
\beq
P_\alpha(\bq) =  (b_\alpha + f\mu_q^2)^2  P^0(q)\,,
\eeq
leading to the well known expressions for the linear monopole and  quadrupole,
\beq
\!\!\!\!\!\!\! \!\!\!\!\!\!\!\!\!\! \!\!\!\!\!\!  \!\!\!\!\!\!  P_\alpha^{(0)}(q) = b_\alpha^2 \left(1+\frac{2}{3} \beta_\alpha +\frac{1}{5}\beta_\alpha^2 \right) \, P^0(q)\,,\quad \frac{P_\alpha^{(2)}(q)}{P_\alpha^{(0)}(q)} = \frac{4 \beta_\alpha }{21} \frac{7+3\beta_\alpha}{1+\frac{2}{3} \beta_\alpha +\frac{1}{5}\beta_\alpha^2}\,,
\label{PSmonquad}
\eeq
where $\beta_\alpha\equiv f/b_\alpha$.

Using the relations above, the RHS of \re{CRRS} can be written as
\beq
 -\frac{P_\alpha^{(0)}(q)}{ \left(1+\frac{2}{3} \beta_\alpha +\frac{1}{5}\beta_\alpha^2 \right)  } \left(\frac{\mu^2}{b_\alpha}  + \beta_\alpha \,\mu \mu_k \mu_q\right) \left(1+ \beta_\alpha\, \mu_q^2\right) \frac{ \partial P_\alpha(\bk)}{\partial \ln{k}} \,.
\eeq

While the definition of the PS multipoles is unique, concerning the BS, different multipoles can be defined, as we can weight the angular integrations with Legendre polynomials in $\mu$, $\mu_k$, or $\mu_q$. From Eq.~\re{murel}, we define BS multipoles with respect to $\mu$ as
\beq
\!\!\!\!\!\! \!\!\!\!\!\! \!\!\!\!\!\! \!\!\!\!\!\! \!\!\!\! B_\alpha^{(l)}(q,k)\equiv (2l+1)\int_{-1}^1\frac{d \mu_k}{2}\int_{-1}^1\frac{d \mu_q}{2} \int_0^{2\pi}\frac{d\phi}{2\pi}\,B_\alpha(\bq,\bk_-,-\bk_+)  \,{\cal P}_l(\mu(\mu_k,\mu_q,\phi))\,.
\eeq
The CR's for the monopole and the quadrupole in $\mu$ then read,
\beqra
&&\!\!\!\!\!\! \!\!\!\! \!\!\!\!\!\! \!\!\!\! \! \lim_{q/k\to 0} \frac{B_\alpha^{(0)}(q,k)}{P_\alpha^{(0)}(q) P_\alpha^{(0)}(k)}= -\left[\frac{1}{3 b_\alpha} +\frac{b_\alpha-1}{9 \,b_\alpha}\beta_\alpha\frac{1+\frac{3}{5}\beta_\alpha}{1+\frac{2}{3} \beta_\alpha+\frac{1}{5}\beta_\alpha^2} \right] \frac{d \ln P_\alpha^{(0)}(k)}{d\ln{k}}\nonumber\\
&&\qquad\quad\quad\;\;\;\;\;\; \;- \frac{2 \beta_\alpha [2 +b_\alpha(5+3\beta_\alpha)]}{225\, b_\alpha \left(1+\frac{2}{3} \beta_\alpha+\frac{1}{5}\beta_\alpha^2\right) } \frac{P_\alpha^{(2)}(k)}{P_\alpha^{(0)}(k)} \frac{d \ln P_\alpha^{(2)}(k)}{d\ln{k}} + \cdots \,,\nonumber\\
&&\!\!\!\!\!\! \!\!\!\! \!\!\!\!\!\! \!\!\!\!\! \lim_{q/k\to 0}\frac{B_\alpha^{(2)}(q,k)}{P_\alpha^{(0)}(q) P_\alpha^{(0)}(k)}= -2 \left[\frac{1}{3 b_\alpha} +\frac{(b_\alpha-1)}{9 \,b_\alpha}\beta_\alpha\frac{1+\frac{3}{5}\beta_\alpha}{1+\frac{2}{3} \beta_\alpha+\frac{1}{5}\beta_\alpha^2} \right] \frac{d \ln P_\alpha^{(0)}(k)}{d\ln{k}}\nonumber\\
&&\qquad\quad\quad\;\;\;\;\;\; \;- \frac{2 \beta_\alpha [77 +b_\alpha(98+75\beta_\alpha)]}{2205\, b_\alpha \left(1+\frac{2}{3} \beta_\alpha+\frac{1}{5}\beta_\alpha^2\right) } \frac{P_\alpha^{(2)}(k)}{P_\alpha^{(0)}(k)}\frac{d \ln P_\alpha^{(2)}(k)}{d\ln{k}}\nonumber\\
&&\qquad\quad\quad\;\;\;\;\;\; \;- \frac{8 \beta_\alpha^2}{735\, \left(1+\frac{2}{3} \beta_\alpha+\frac{1}{5}\beta_\alpha^2\right) }\frac{P_\alpha^{(4)}(k)}{P_\alpha^{(0)}(k)}\frac{d \ln P_\alpha^{(4)}(k)}{d\ln{k}} +\cdots\,,
\label{takaCR}
\eeqra
where dots indicate smooth/subdominant contributions.
Taking multipoles with respect to $\mu_k$ and $\mu_q$, defined as
\beq
\!\!\!\!\!\! \!\!\!\! \!\!\!\!\!\! \!\!\!\!\!B^{(l_{k,q})}_{\alpha}(q,k)\equiv \frac{2 l_{k,q} + 1}{2}\int_{-1}^1\frac{d\mu_k}{2}\int_{-1}^1\frac{d\mu_q}{2}\int^{2\pi}_0\frac{d\phi}{2\pi} B_{\alpha}(\bq,\bk_-.- \bk_+)\mathcal{P}_{l_{k,q}}(\mu_{k,q}),
\eeq
we get the same monopole equation as above, while, for the quadrupoles, we get,
\beqra
&&\!\!\!\!\!\! \!\!\!\! \!\!\!\!\!\! \!\!\!\!\! \lim_{q/k\to 0}\frac{B_\alpha^{(l_k=2)}(q,k)}{P_\alpha^{(0)}(q) P_\alpha^{(0)}(k)}= -\frac{2\beta_\alpha}{45 \,b_\alpha}\frac{2+b_\alpha(5 + 3\beta_\alpha)}{1+\frac{2}{3} \beta_\alpha+\frac{1}{5}\beta_\alpha^2} \frac{d \ln P_\alpha^{(0)}(k)}{d\ln{k}}\nonumber\\
&&\qquad\quad\quad\;\;\;\;\;\; \;- \frac{ 105  +43 \beta_\alpha + 55 b_\alpha \beta_\alpha + 33 b_\alpha\beta_\alpha^2}{315\, b_\alpha \left(1+\frac{2}{3} \beta_\alpha+\frac{1}{5}\beta_\alpha^2\right) } \frac{P_\alpha^{(2)}(k)}{P_\alpha^{(0)}(k)}\frac{d \ln P_\alpha^{(2)}(k)}{d\ln{k}}\nonumber\\
&&\qquad\quad\quad\;\;\;\;\;\; \;- \frac{4 \beta_\alpha}{315 b_\alpha}\frac{2 + b_\alpha (5 + 3\beta_\alpha)}{1+\frac{2}{3} \beta_\alpha+\frac{1}{5}\beta_\alpha^2}\frac{P_\alpha^{(4)}(k)}{P_\alpha^{(0)}(k)}\frac{d \ln P_\alpha^{(4)}(k)}{d\ln{k}} +\cdots\,,
\label{takaCRmuk}
\eeqra
and
\beqra
&&\!\!\!\!\!\! \!\!\!\! \!\!\!\!\!\! \!\!\!\!\! \lim_{q/k\to 0}\frac{B_\alpha^{(l_q=2)}(q,k)}{P_\alpha^{(0)}(q) P_\alpha^{(0)}(k)}= -\frac{2\beta_\alpha}{63 \,b_\alpha}\frac{7+b_\alpha(7 + 6\beta_\alpha)}{1+\frac{2}{3} \beta_\alpha+\frac{1}{5}\beta_\alpha^2} \frac{d \ln P_\alpha^{(0)}(k)}{d\ln{k}}\\
&&\qquad\quad\quad\;\;\;\;\;\; \;- \frac{ 42  +22 \beta_\alpha + 28 b_\alpha \beta_\alpha + 24 b_\alpha\beta_\alpha^2}{315\, b_\alpha \left(1+\frac{2}{3} \beta_\alpha+\frac{1}{5}\beta_\alpha^2\right) } \frac{P_\alpha^{(2)}(k)}{P_\alpha^{(0)}(k)}\frac{d \ln P_\alpha^{(2)}(k)}{d\ln{k}} +\cdots\,\nonumber.
\label{takaCRmuq}
\eeqra
Notice that, unlike the quadrupole in $\mu$, those in $\mu_k$ and $\mu_q$ are proportional to $\beta_\alpha$, and therefore are non-vanishing only in redshift space.

\section{Simulations}
\label{sims}
We analyse the same set of simulations already presented in Ref.~\cite{Marinucci:2019wdb}. The trajectories of $2048^3$ particles are followed by a public Tree-Particle Mesh code, \textsc{Gadget2}~\cite{Springel:2005mi}, in periodic comoving boxes with $(4\,h^{-1}\mathrm{Gpc})^3$ assuming a flat-$\Lambda$CDM cosmology consistent with the Planck satellite~\cite{Ade:2015xua}. The initial particle displacements as well as the velocities are set up with a second-order Lagrangian perturbation theory (2LPT; \cite{Scoccimarro:1997gr,Crocce:2006ve}) code implemented initially in Ref.~\cite{Nishimichi:2008ry} and then parallelized in Ref.~\cite{Valageas:2010yw}. The other simulation parameters can be found in Ref.~\cite{Nishimichi:2018etk}. We newly performed ten random realizations for this project in larger simulation boxes compared to those presented in Ref.~\cite{Nishimichi:2018etk}, which is either $(1\,h^{-1}\mathrm{Gpc})^3$ or $(2\,h^{-1}\mathrm{Gpc})^3$, to examine the squeezed-limit of the BS more precisely. We store the particle snapshots at $z=0$ and $1$. Dark matter halos are identified at these redshifts with a phase-space based finder, \textsc{Rockstar}~\cite{Behroozi:2011ju}.

We measure the PS and the BS using fast Fourier transform. We first assign the particle mass or the halo number density on to $1024^3$ grid points using Cloud-in-Cells (CIC) algorithm \cite{hockney81} in configuration space. After transforming to the Fourier space, we mitigate the aliasing effect~\cite{Jing:2004fq} using the interlacing technique~\cite{Sefusatti:2015aex} and then divide the field by the CIC window function. We store the products of the resulting fields into bins to form the estimator of either the PS and the BS. In case of the PS, we prepare bins with the interval of $0.005\,\hmp$. This is sufficient to resolve the BAO feature in detail. The product, $|\delta_\mathbf{k}|^2$, is averaged in the bins to obtain our estimator of the PS. In redshift space, we also consider $\mathcal{P}_\ell(\mu_\mathbf{k})|\delta_\mathbf{k}|^2$ to estimate the multipole moments. In case of the halo PS, we subtract the standard Poissonian shot noise contribution, $V/N_\mathrm{h}$, where $V$ is the simulation volume and $N_\mathrm{h}$ is the number of halos, from the monopole moment.

The estimator of the BS can be constructed in an analogous manner. We refined the binning scheme from that adopted in Ref.~\cite{Marinucci:2019wdb} to better capture  its configuration dependence. We consider a pair of wavevectors $(\mathbf{k}, \mathbf{q})$ and form a triangle $(\mathbf{q},-\mathbf{k_+},\mathbf{k_-})$. We bin the triangles in $q$ and $k$  at every $0.01\,\hmp$, and then we sum the ratio of the bispectrum to the linear PS in $q$ up to a given $\qm$, weighting the sum with the number of triangles in each $q$-bin,

\beq
\sum_{q\le \qm} N_{\rm tri}(q,k_i)\frac{B_\alpha^{(l)}(q,k_i)}{P_\alpha^{(0)}(q)}\equiv \frac{B_\alpha^{(l)}}{P_\alpha^{(0)} }(q_{\rm max},k_i)\,.
\label{qmax}
\eeq
 The remaining degree of freedom, the angle between the two wavevectors (and also the relative angle with respect to the line-of-sight direction in case of redshift space), is integrated to obtain the moment estimators: in Ref.~\cite{Marinucci:2019wdb}, we instead kept the angle dependence and estimated the BS in bins of $q, k$ and $\mu$. Since we know the expected angle dependence of the oscillatory feature, that is simply $\mu^2$, we can fully express it with the first two even moments, monopole ($\ell=0$) and the quadrupole ($\ell=2$). This helps to obtain the BAO feature with smaller error bars.
We subtracted the shot noise, $(V/N_\mathrm{h})^2+(V/N_\mathrm{h})[P_\mathrm{h}(q)+P_\mathrm{h}(k_+)+P_\mathrm{h}(k_-)]$ from the halo monopole BS.

\section{Results}
\label{resid}

In this section we describe our procedure to evaluate the bias $b_\alpha$ and the parameter $\beta_{\alpha}$ from the simulations, by using the CR's.

In order to do that, we have to fit, in $k$,  the LHS's of the CR, binned up to a given $\qm$, see Eq.~\re{qmax},
\beq
  \frac{1}{P_\alpha^{(0)}(k_i)}\frac{B_\alpha^{(l)}}{P_\alpha^{(0)} }(q_{\rm max},k_i)\,,
 \label{LHCR}
\eeq
with the RHS's, which we will model as
\beqra
&&-  \sum_{l'=0,2}C^{(l,l')}(b_\alpha,\beta_\alpha) \left( \frac{d \ln P_\alpha^{(l')}(k_i)}{d\ln{k}}- \frac{d \ln P_\alpha^{(l')}(k_i)}{d\ln{k}} \Bigg|_{\rm smooth}  \right)e^{-c(q_{max}) k^2}
\nonumber\\
&&\qquad+ p(\{a^{(l)}_i(\qm)\};k_i)\,,
 \label{RHCR}
\eeqra
where the coefficients $ C^{(l,l')}(b_\alpha,\beta_\alpha)$ can be read from Eq.~\re{takaCR}, while the smooth functions $p(\{a^{(l)}_i(q)\};k)$ are going to fit the  smooth contributions from the derivatives of the PS multipoles  together with the other smooth and or subdominant contributions discussed in Sect.~\ref{CRdisc}. The contribution of the PS hexadecapole ($l=4$) to the second of Eqs.~\re{takaCR}, is numerically negligible, and we do not include it in our analysis.

We have isolated the smooth contributions from the derivatives of the PS monopole and quadrupole by subtracting a spline 
fit.
We tested alternative algorithms to extract the smooth contributions obtaining stable results for the extracted parameters. Moreover, we introduced a scale dependent BAO damping term,  $e^{-c(q_{max}) k^2}$, which models possible correlations  between the long mode and the FoG damping beyond the squeezed limit.


\begin{figure}[htbp]
\centering\includegraphics[width=5cm]{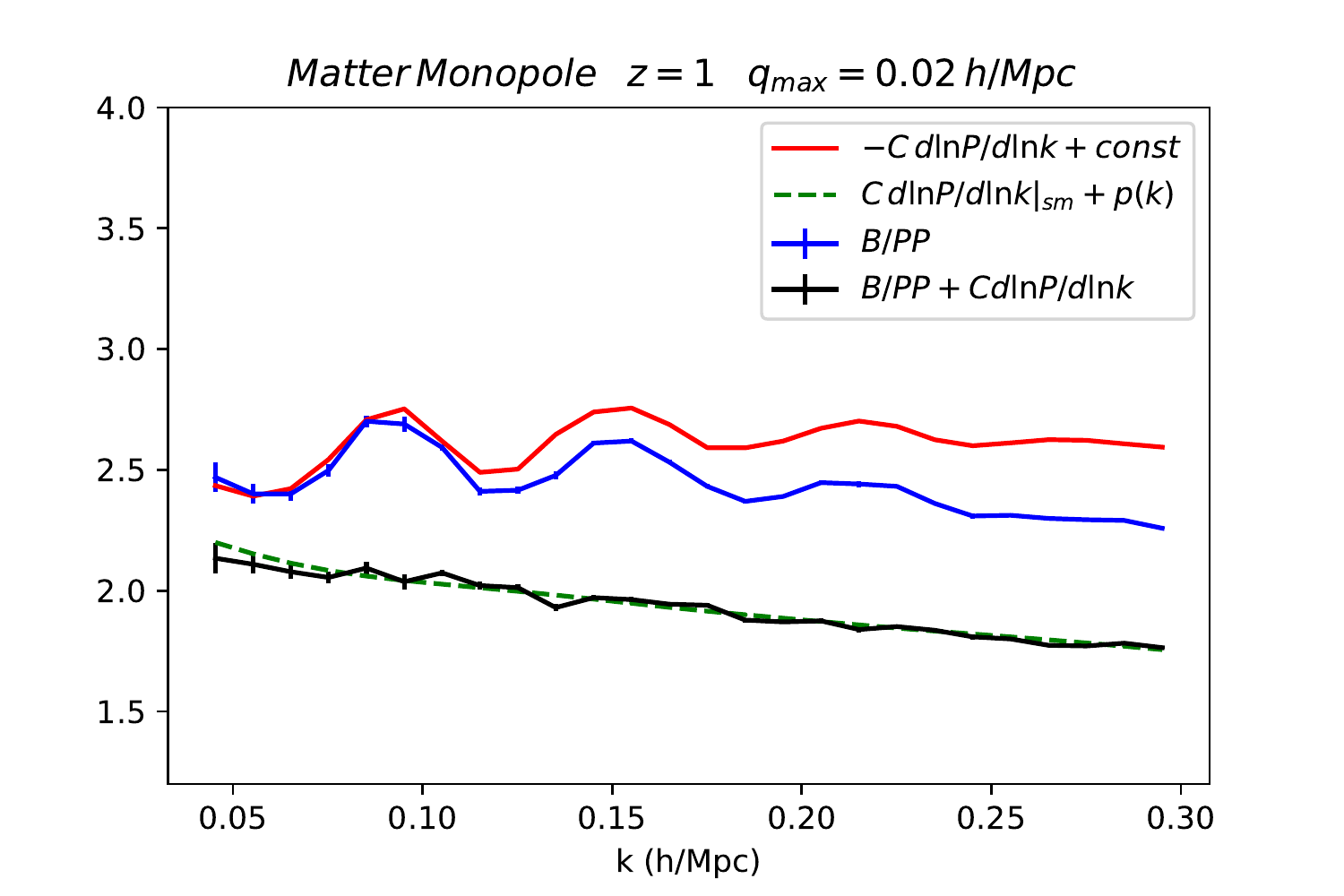}
\centering\includegraphics[width=5cm]{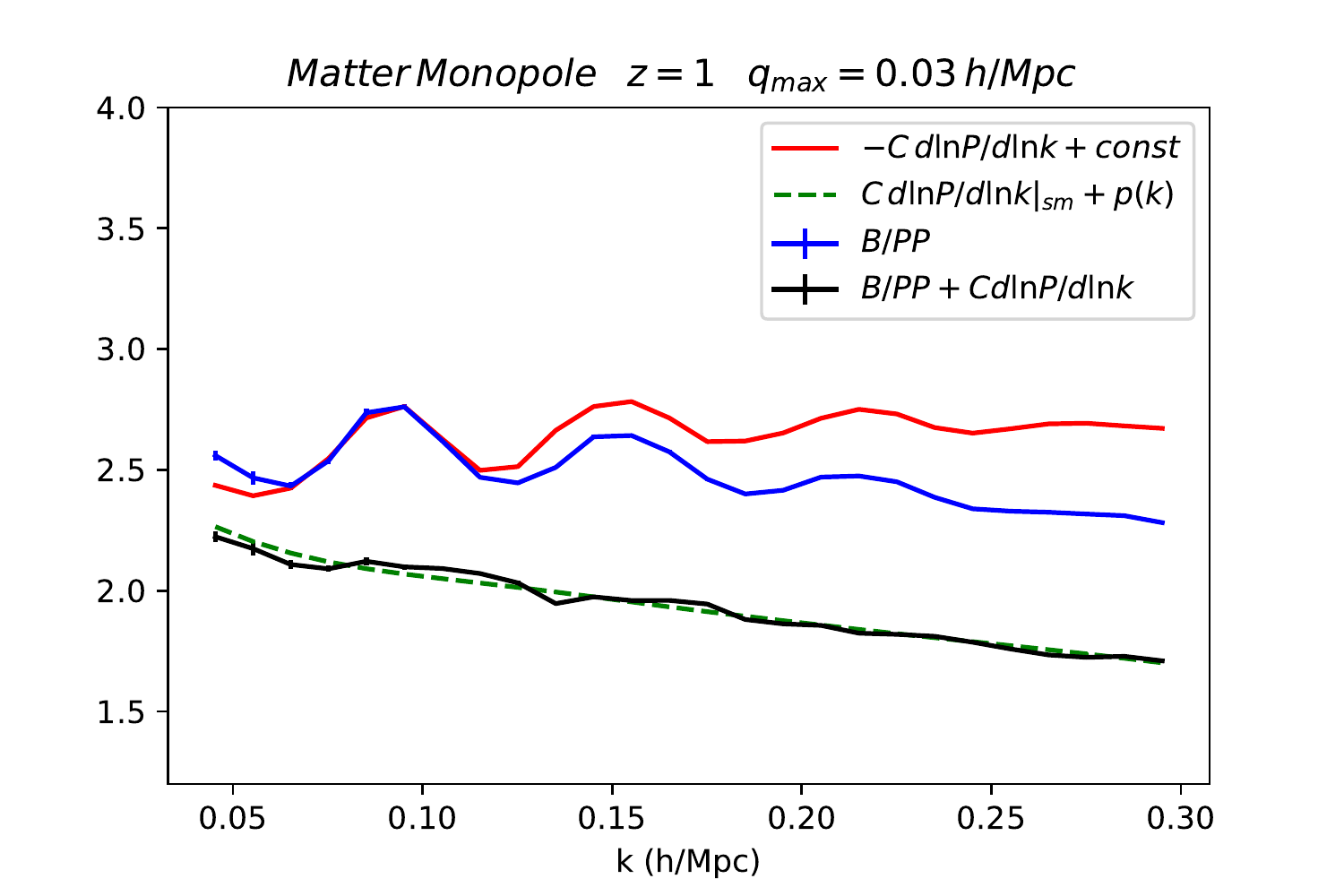}
\centering\includegraphics[width=5cm]{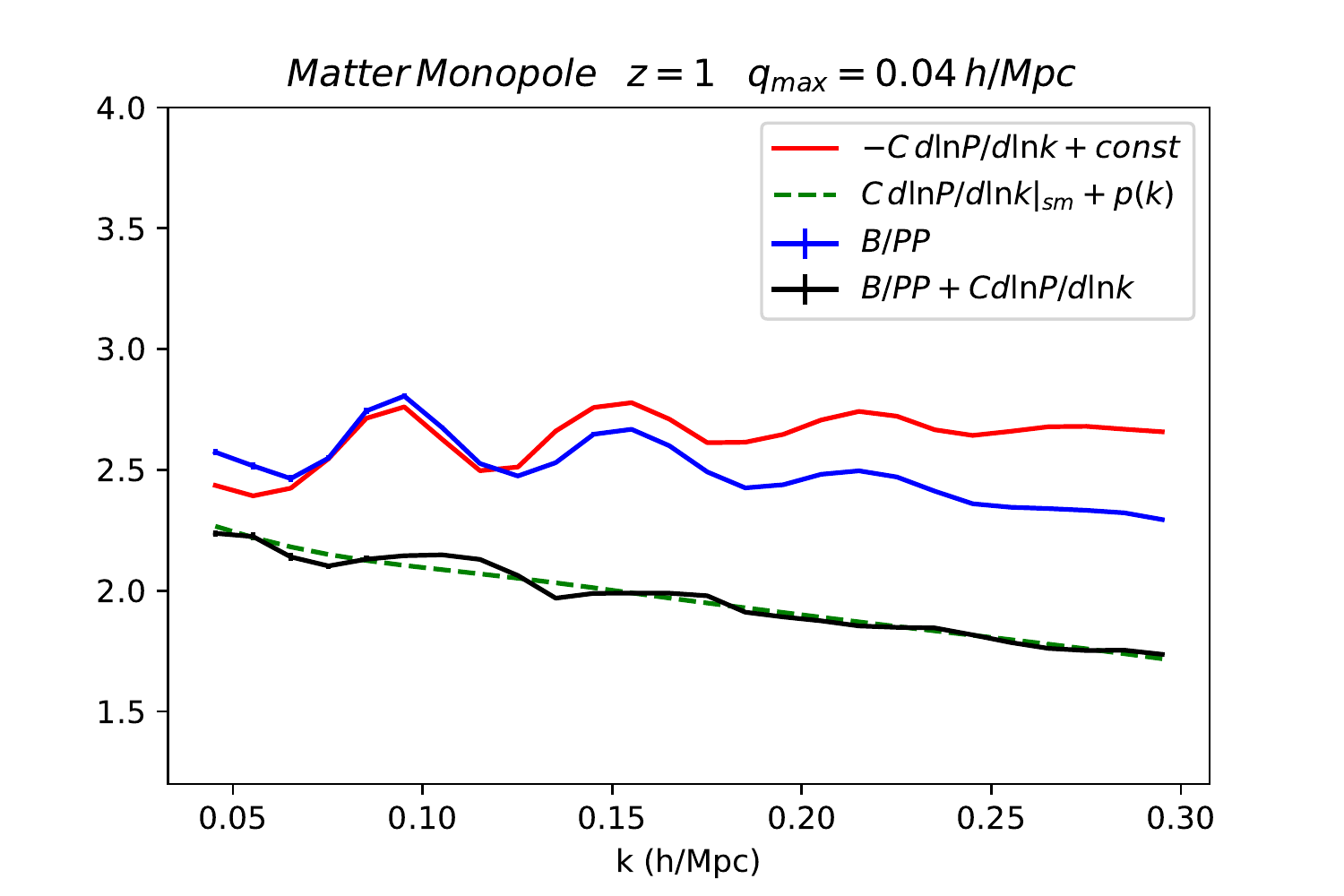}
\centering\includegraphics[width=5cm]{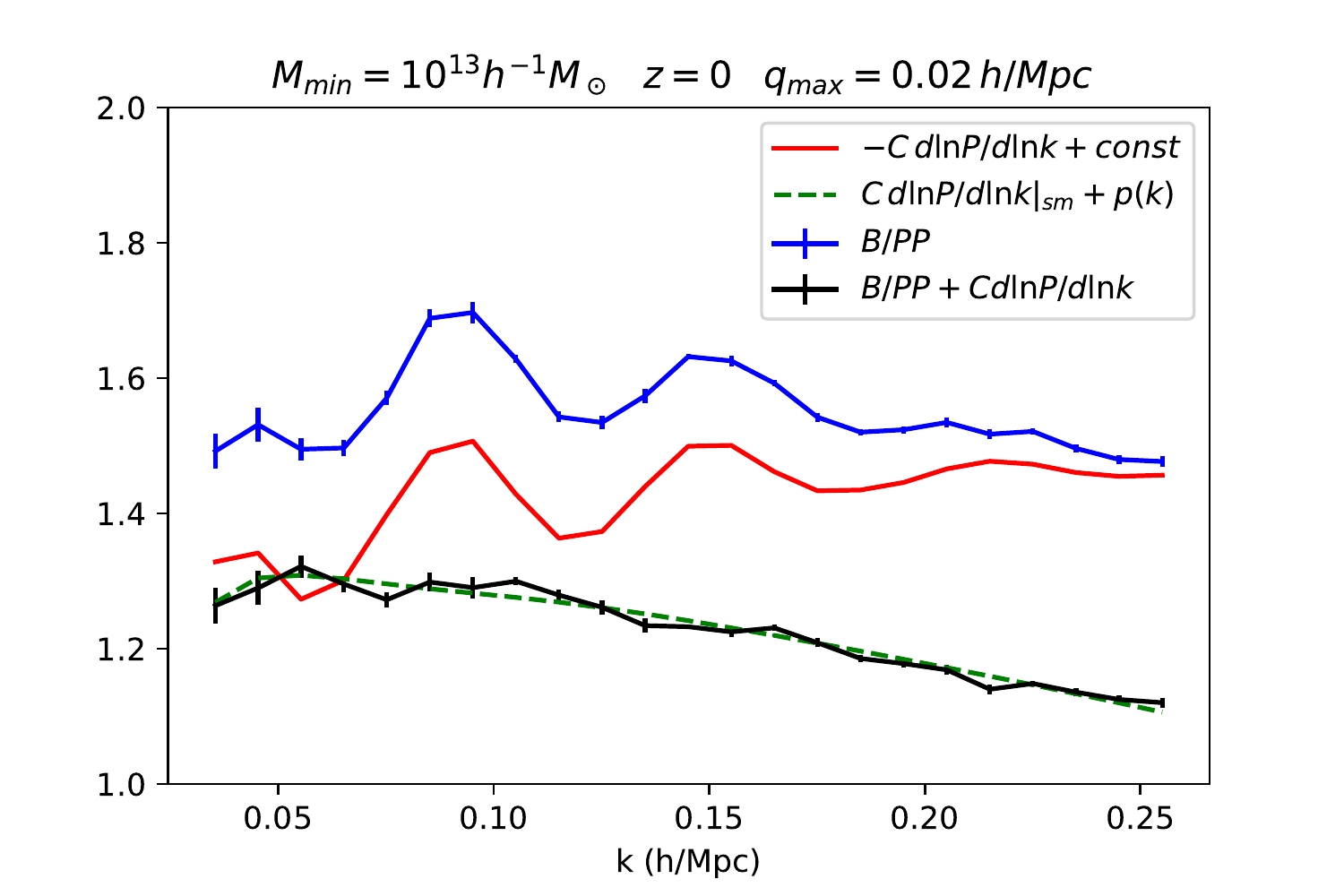}
\centering\includegraphics[width=5cm]{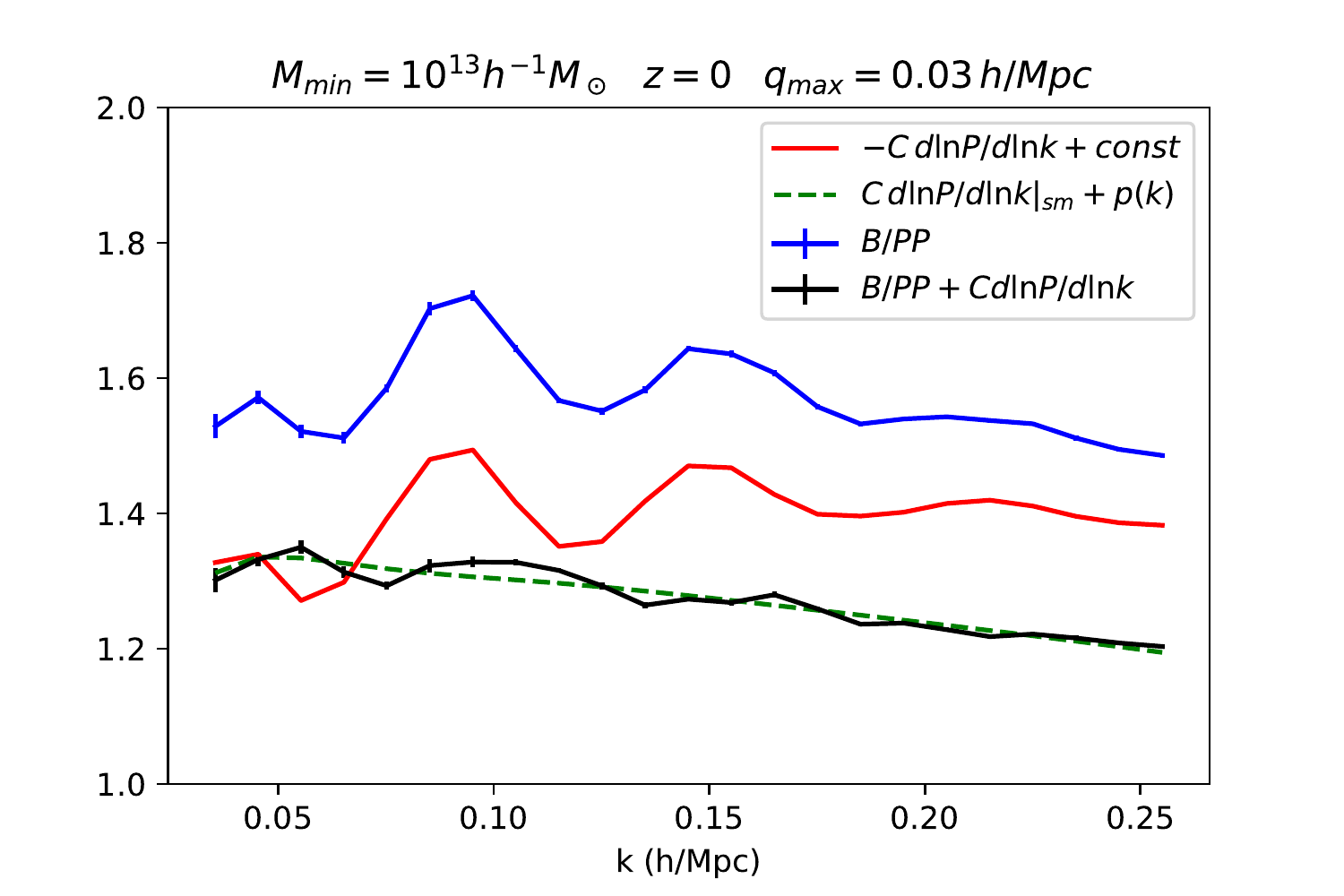}
\centering\includegraphics[width=5cm]{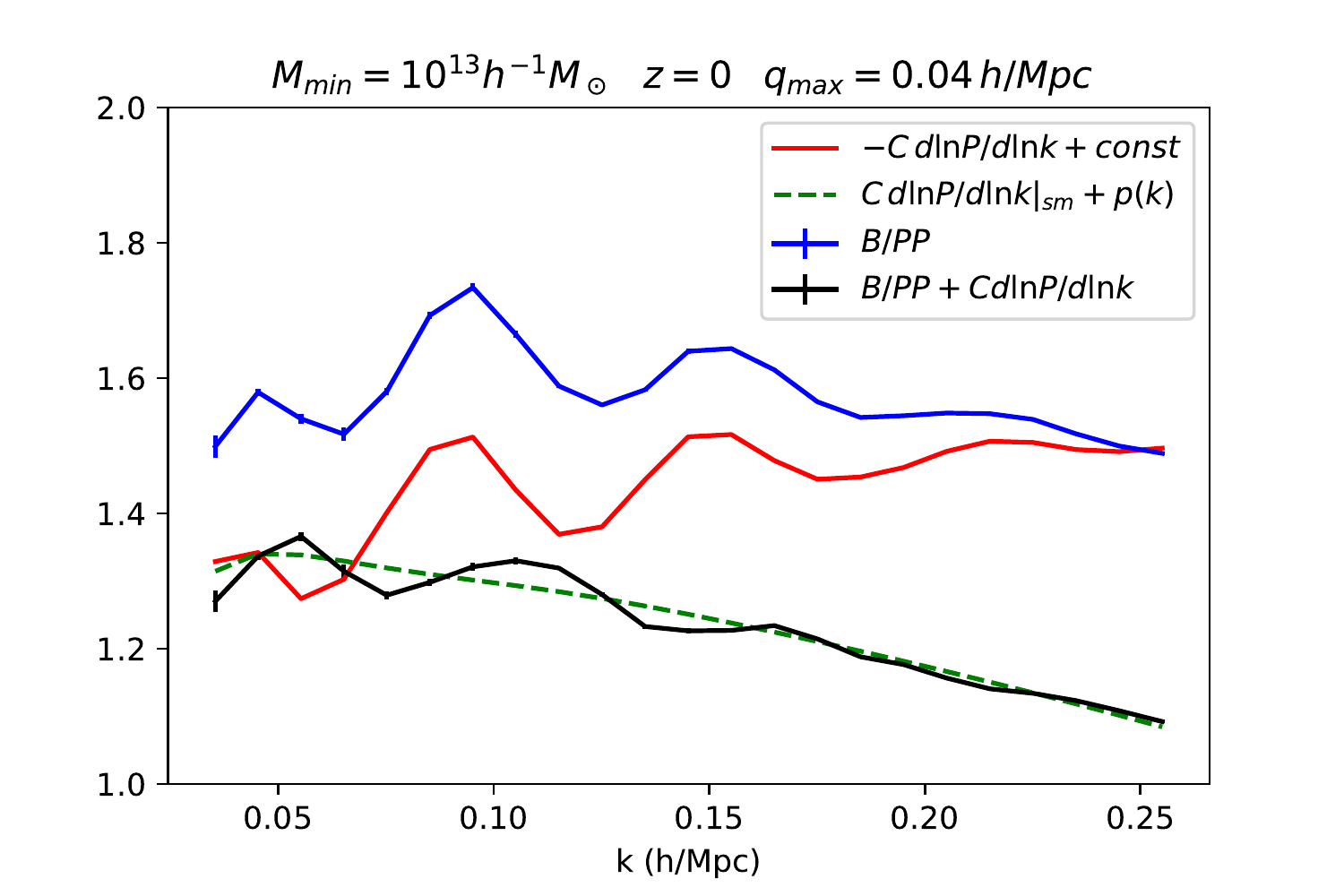}
\caption{The BS to PS's ratios of Eq.~\re{LHCR} (blue lines with error bars), and the oscillating part of the terms containing the logarithmic derivatives of the PS's in Eq.~\re{RHCR} (red lines). The difference between the two is given by the black lines, together with the smooth fitting functions described in the text (green-dashed lines). The red lines include a constant offset for graphical purposes. The fiducial values for $b_\alpha$, $\beta_\alpha$ and the best fitting values for the nuisance parameters have been used to produce these plots.}
\label{residuals}
\end{figure}

\begin{figure}[htbp]
\centering\includegraphics[width=5cm]{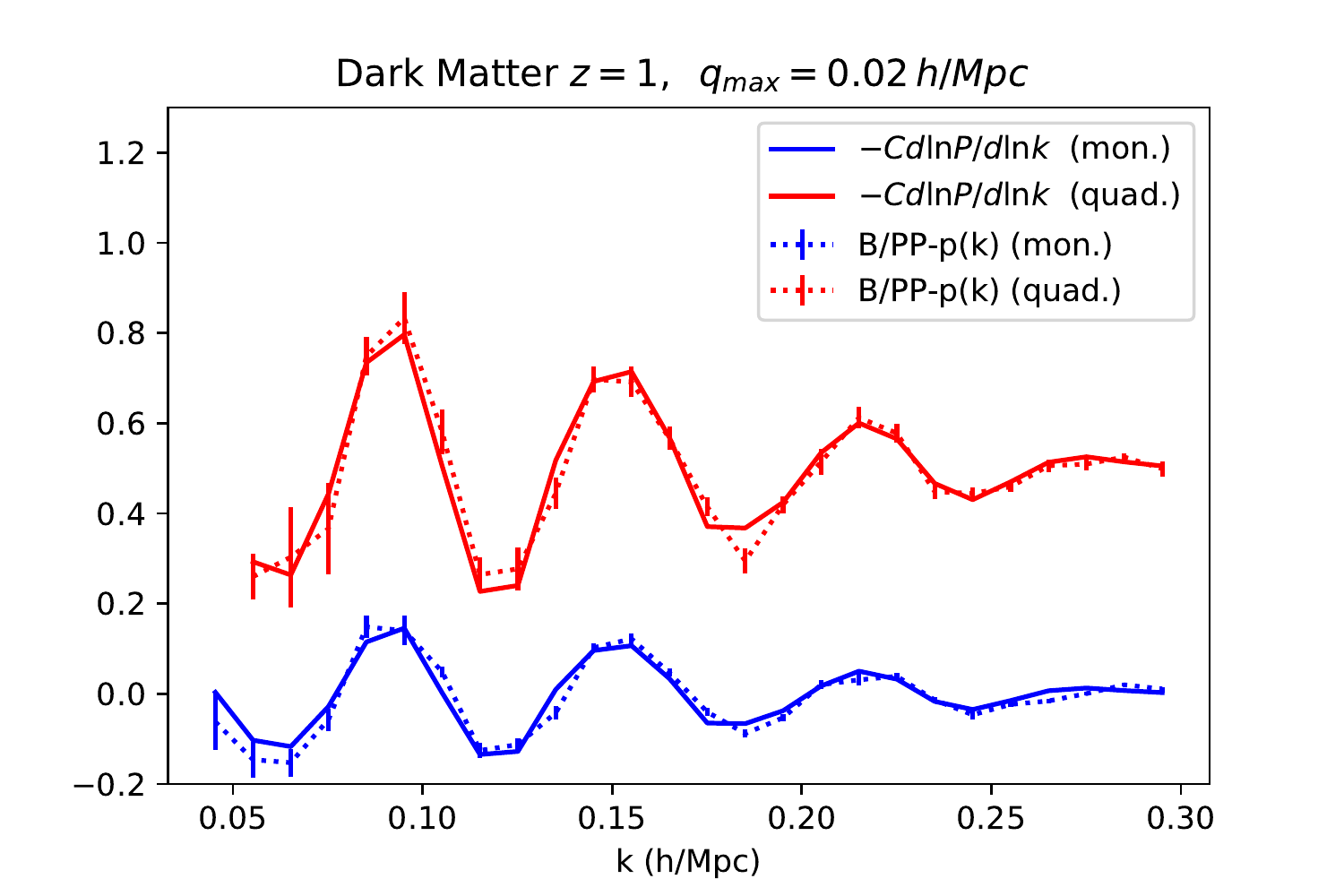}
\centering\includegraphics[width=5cm]{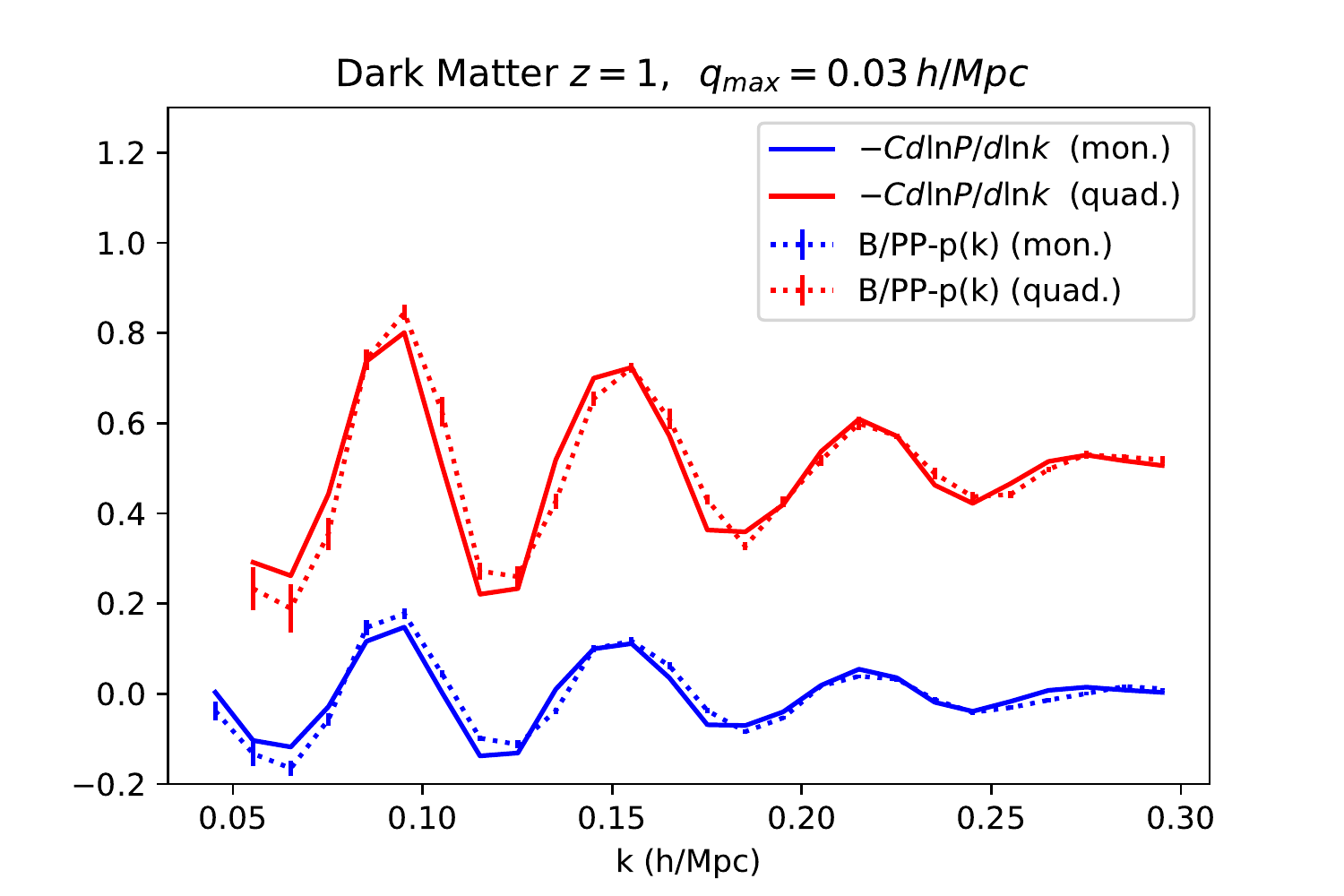}
\centering\includegraphics[width=5cm]{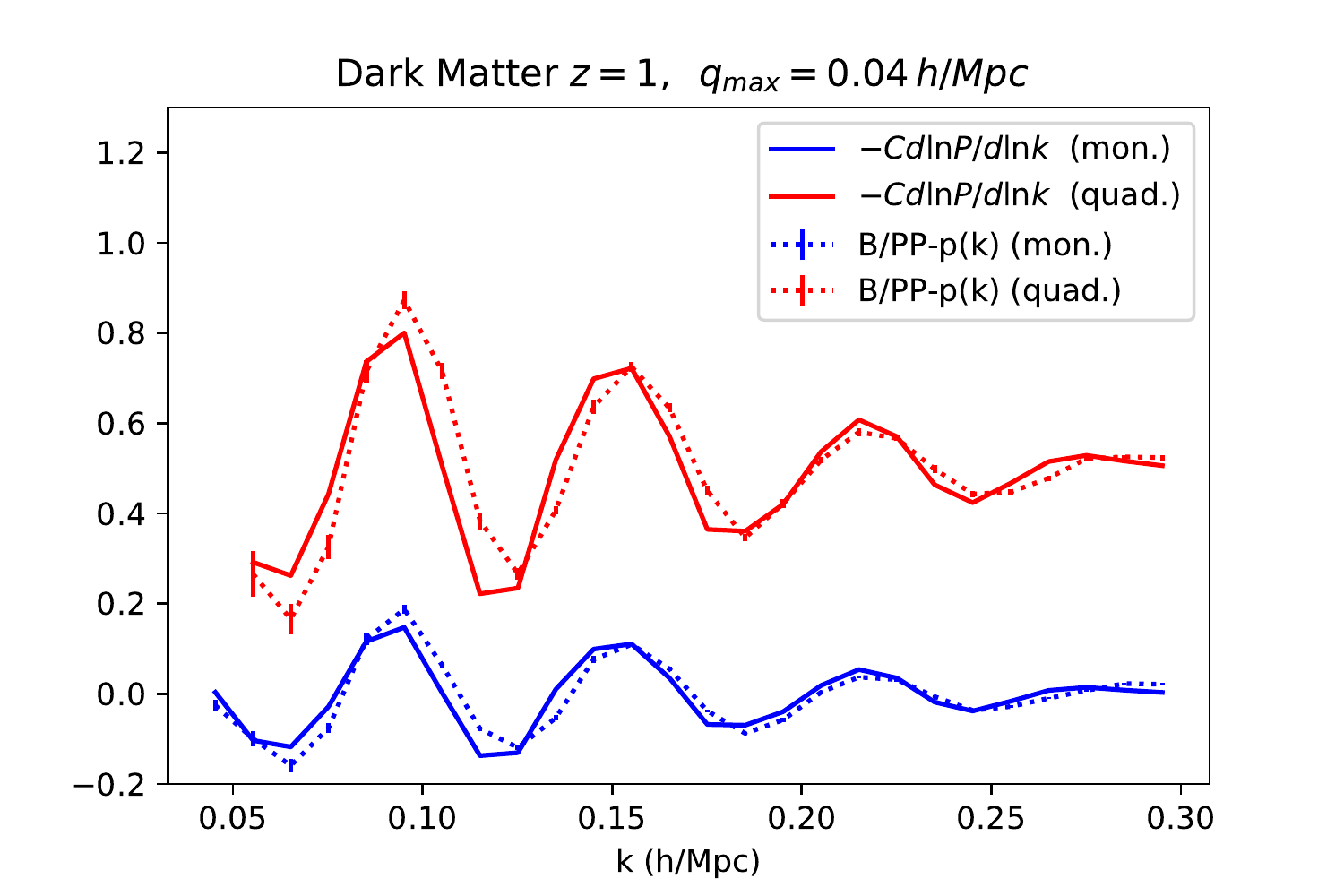}
\centering\includegraphics[width=5cm]{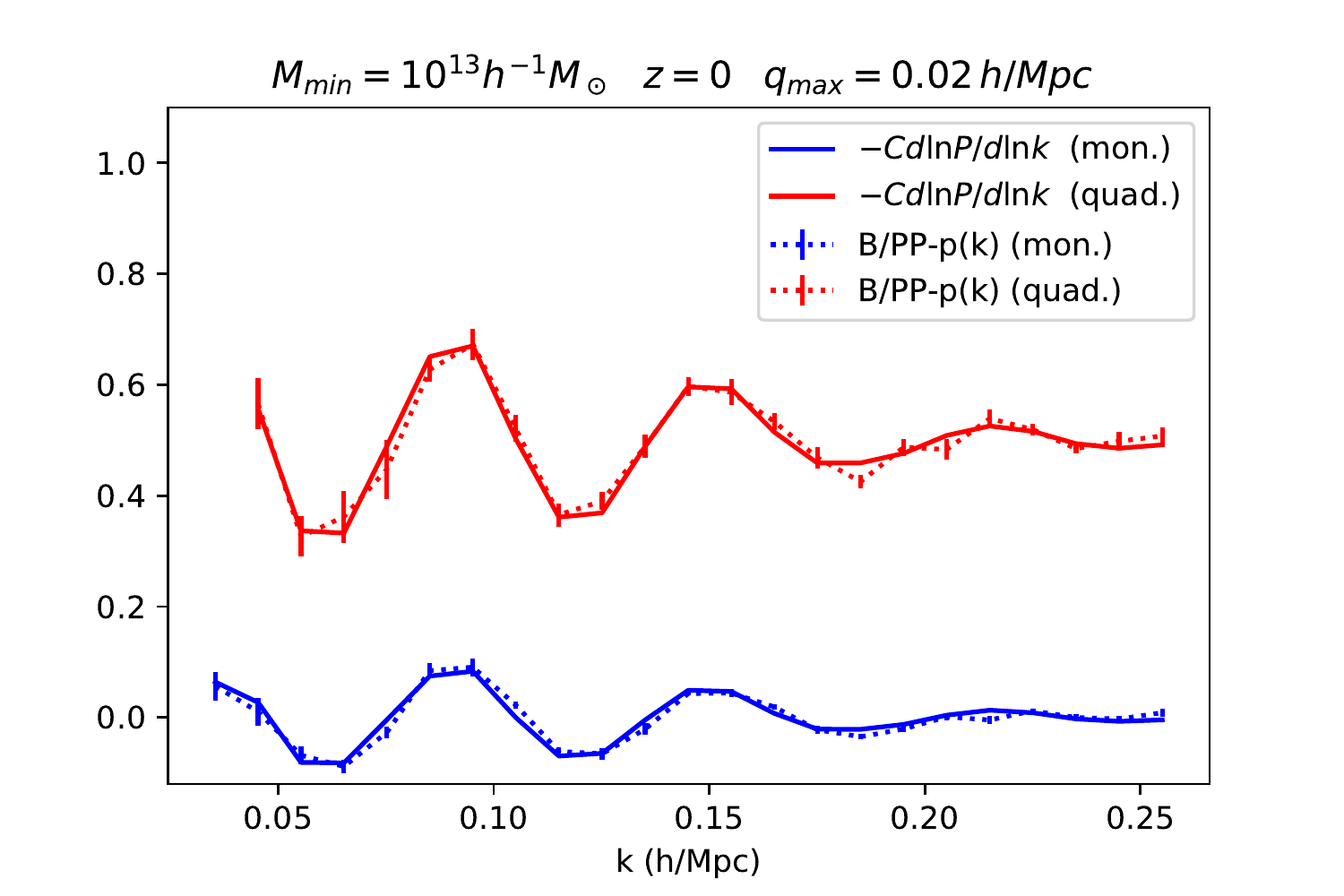}
\centering\includegraphics[width=5cm]{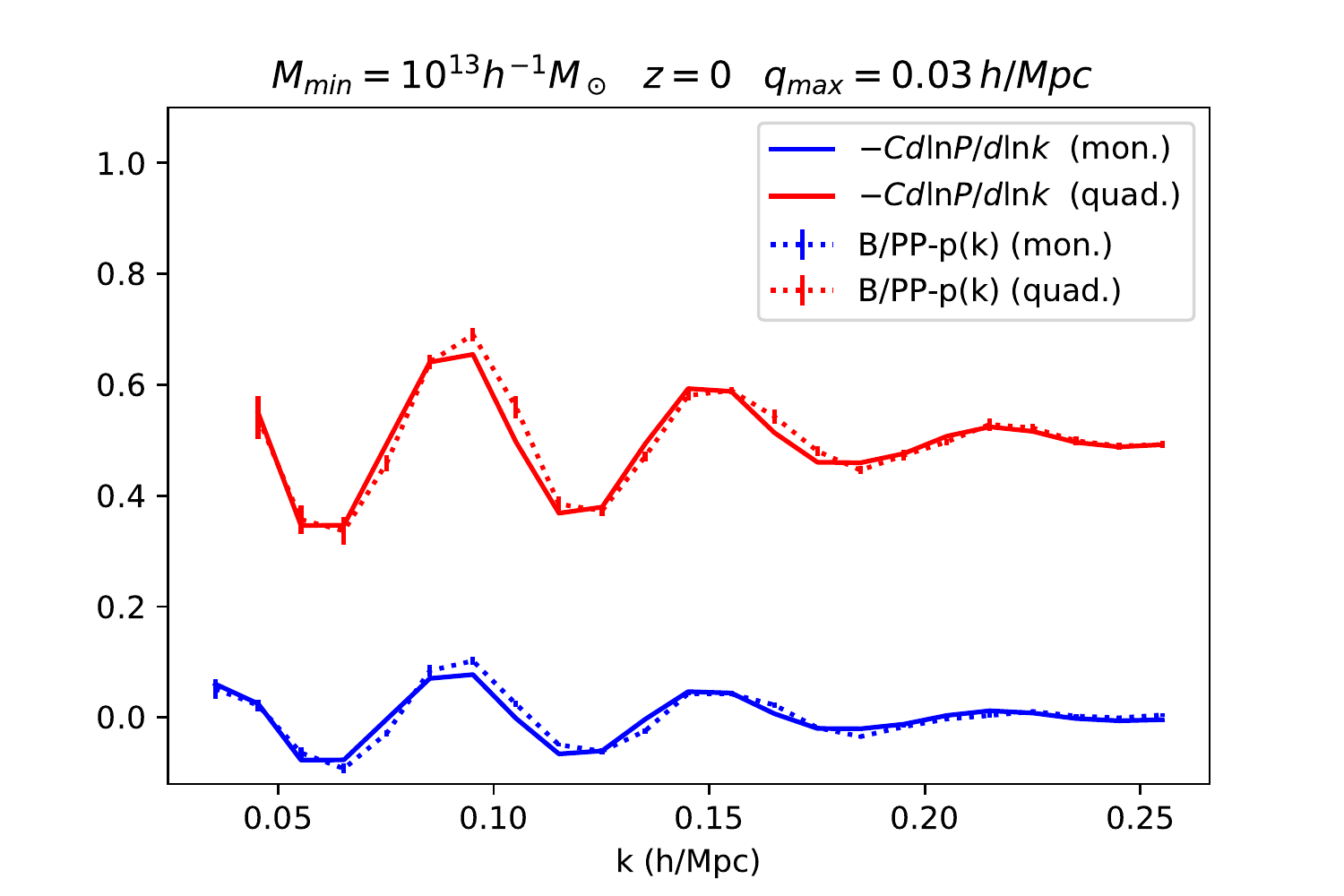}
\centering\includegraphics[width=5cm]{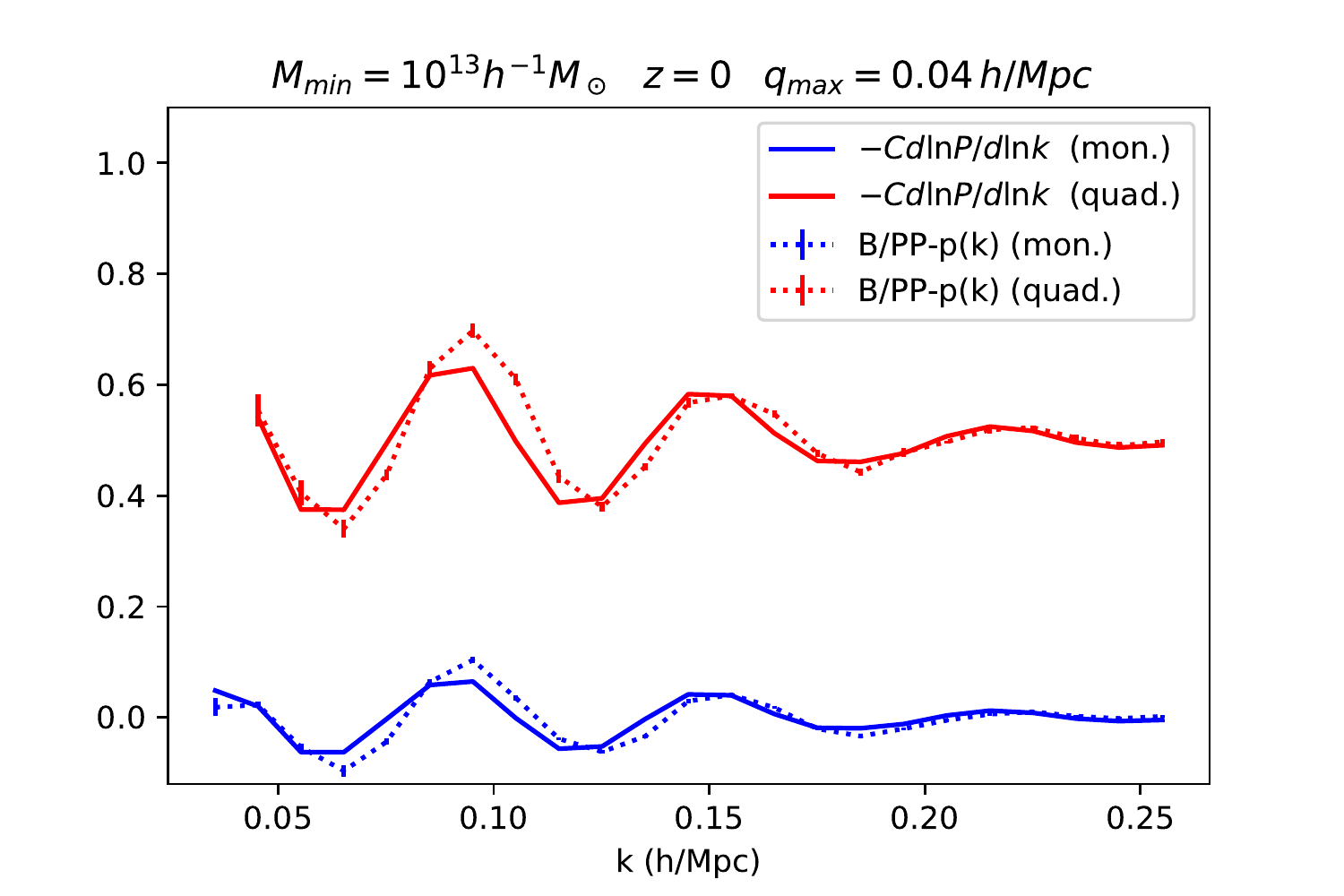}
\caption{Comparison between the oscillating parts of the BS to PS ratios in Eq.~\re{LHCR} (dotted lines) and that of Eq.~\re{RHCR} (solid lines) both for the monopole (blue) and the quadrupole (red) components of the BS. The fiducial values for $b_\alpha$, $\beta_\alpha$ and the best fitting values for the nuisance parameters have been used to produce these plots.}
\label{fits}
\end{figure}


The form of the fitting functions $p(\{a^{(l)}_i(q)\};k)$ is chosen in order to reproduce the leading expected contributions. It  contains a term constant in $k$, as the lowest order PT result in the squeezed limit. Then, we include a negative contribution proportional to $k^2$, accounting for the leading contribution to the logarithmic derivative of the nonlinear PS from the Fingers of God effect. This can be understood by looking at the pre-factor in Eq.~\re{fog}. Finally, we  include also
 a $k^{-2}$ term to control possible (small) deviations from the squeezed limit, which we expect to scale as $(\qm/k)^2$.

Summarizing, the smooth function we will use in the CR's for the monopole and the quadrupole of the BS takes the form
\beq
p(\{a^{(l)}_i(\qm)\};k)= a^{(l)}_{-2}(\qm)\left(\frac{k}{\qm}\right)^{-2} + a^{(l)}_0(\qm) + a^{(l)}_2(\qm) \left(\frac{k}{\bar k}\right)^{2},
\label{smoothing}
\eeq
where we have fixed the pivot scale  $\bar k =0.06\,\hmp $.
We have also considered extended polynomial fitting formulas, obtaining consistent results for the parameter estimations.

The smoothness of the difference between the LHS and the first term at the RHS of the CR can be verified from simulations, as we show in Fig.~\ref{residuals}, where we  plot the ratio \re{LHCR} of the BS to the PS's (blue lines), the sum of the logarithmic derivative of the PS multiplied by the appropriate coefficients, as in \re{RHCR}, without the subtraction of the smooth part (red lines), the difference between the two curves (black lines) and the smooth interpolation used to fit the latter, given by the sum of the terms containing the spline fits to the logarithmic derivatives of the PS's and the polynomial $p(\{a^{(l)}_i(\qm)\};k_i$ (green-dashed lines). The  fiducial values of $b_h$ and $f$ have been used to evaluate the coefficients in these curves.  The error bars in these plots are dominated by those of the BS. As we can see, the difference between the BS to PS ratios in \re{LHCR} and the terms in the logarithmic derivatives of the PS is smooth, with residual oscillations increasing as one moves away from the squeezed limit, by increasing  $\qm$. On the other hand, by increasing $\qm$ the statistical errors are reduced, as more triangle configurations contribute to the BS, so a compromise has to be found between statistical power and the goodness of the squeezed limit approximation.

In Fig.~\ref{fits} we show the oscillating components of the BS to PS ratios, both for monopoles (dotted blue) and quadrupoles (dotted red) compared to the oscillating parts of Eq.~\re{RHCR}.

We introduce the following Log-likelihood function,
\beq
\chi^2_{CR}(b_\alpha,\beta_\alpha,\{a^{(l)}_i(q)\},\qm) \equiv \sum_i \left(\frac{r_{(0)}^2(k_i)}{\sigma_{(0),i}^2}+\frac{r_{(2)}^2(k_i)}{\sigma_{(2),i}^2}\right)\,,
\label{chiCR}
\eeq
where the $r_{(l)}(k_i)$'s are the differences between Eq.~\re{RHCR} and  Eq.~\re{LHCR}, and $\sigma_{(l) ,i}$'s are the corresponding errors on the BS measured from the simulations, evaluated in the $i$'th $k$-bin. We neglect the error on the PS as it is much smaller than that on the BS, and we assume diagonal covariances.

\begin{figure}[htbp]
\center{\includegraphics[width=0.45\textwidth]{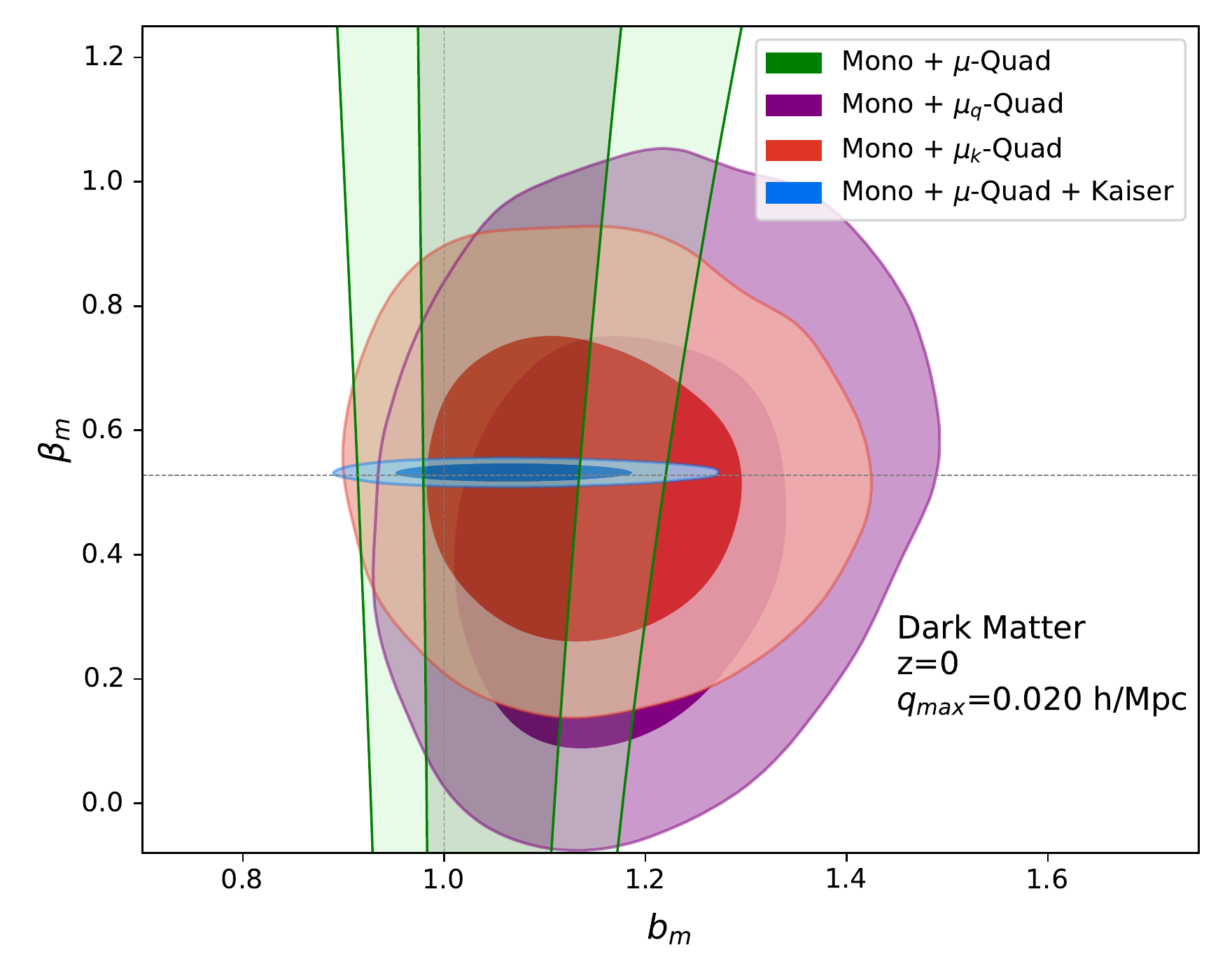}
\includegraphics[width=0.45\textwidth]{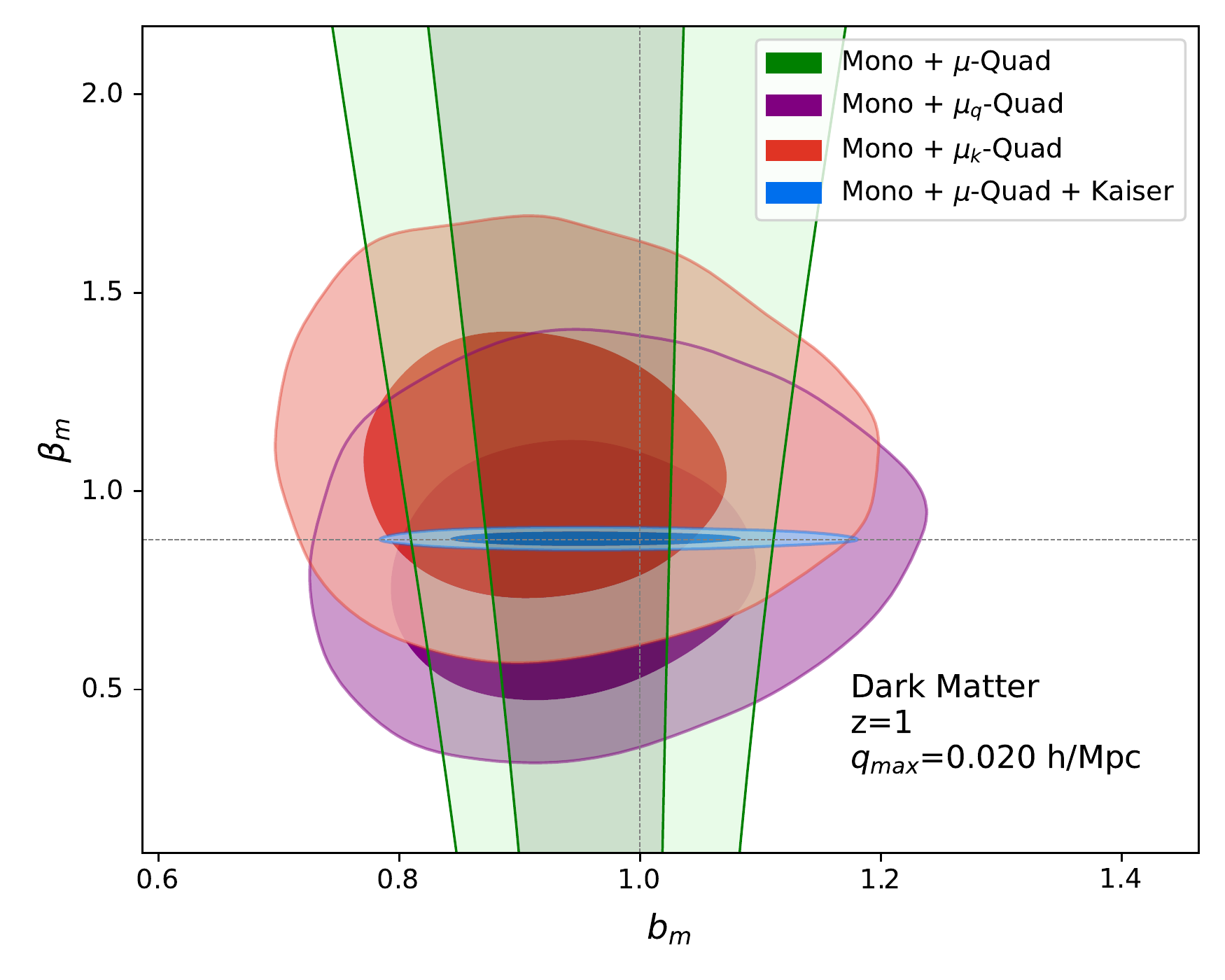}}
\caption{$1$ and $2-\sigma$ constraints in the $b_m-\beta_m$ plane from CR's on the matter BS at redshift $z=0$, (left) and $z=1$ (right). The green contours are obtained combining  the monopole and the $\mu$-quadrupole of the BS (Eq.~\re{takaCR}). Combining the monopole with the BS quadrupole in $\mu_k$ (Eq.~\re{takaCRmuk}), or  the one in $\mu_q$ (Eq.~\re{takaCRmuq}), constrains also $\beta_m$ as shown by the red and purple contours, respectively.  Adding information on the PS quadrupole to monopole ratio, Eq.~\re{rk}, gives the blue contours. The dotted lines indicate the fiducial values for $b_m$ and $\beta_m$.}
\label{fig:CRonly}
\end{figure}

The CR's in Eqs.~\re{takaCR} and \re{takaCRmuk}, depend both on $b_\alpha$ and $\beta_\alpha$, so, in principle, one can break the degeneracy between these two parameters by using the CR's alone.  This is indeed the case, as we show in Fig.~\ref{fig:CRonly}. The BS monopole, as the $\mu-$ quadrupole  (from  Eqs.~\re{takaCR}) are mostly sensitive to $b_\alpha$ but insensitive to $\beta_\alpha$. On the other hand, when we combine the monopole with the $\mu_k$ or $\mu_q$ quadrupoles of Eqs.~\re{takaCRmuk} or ~\re{takaCRmuq} we can constrain also $\beta_\alpha$, although only at the $\sim40\,\%$ level. We do not combine different bispectrum quadrupoles, as they are not independent, and their cross-covariance would be non-trivial.

\begin{figure}[htbp]
\center{\includegraphics[width=0.5\textwidth]{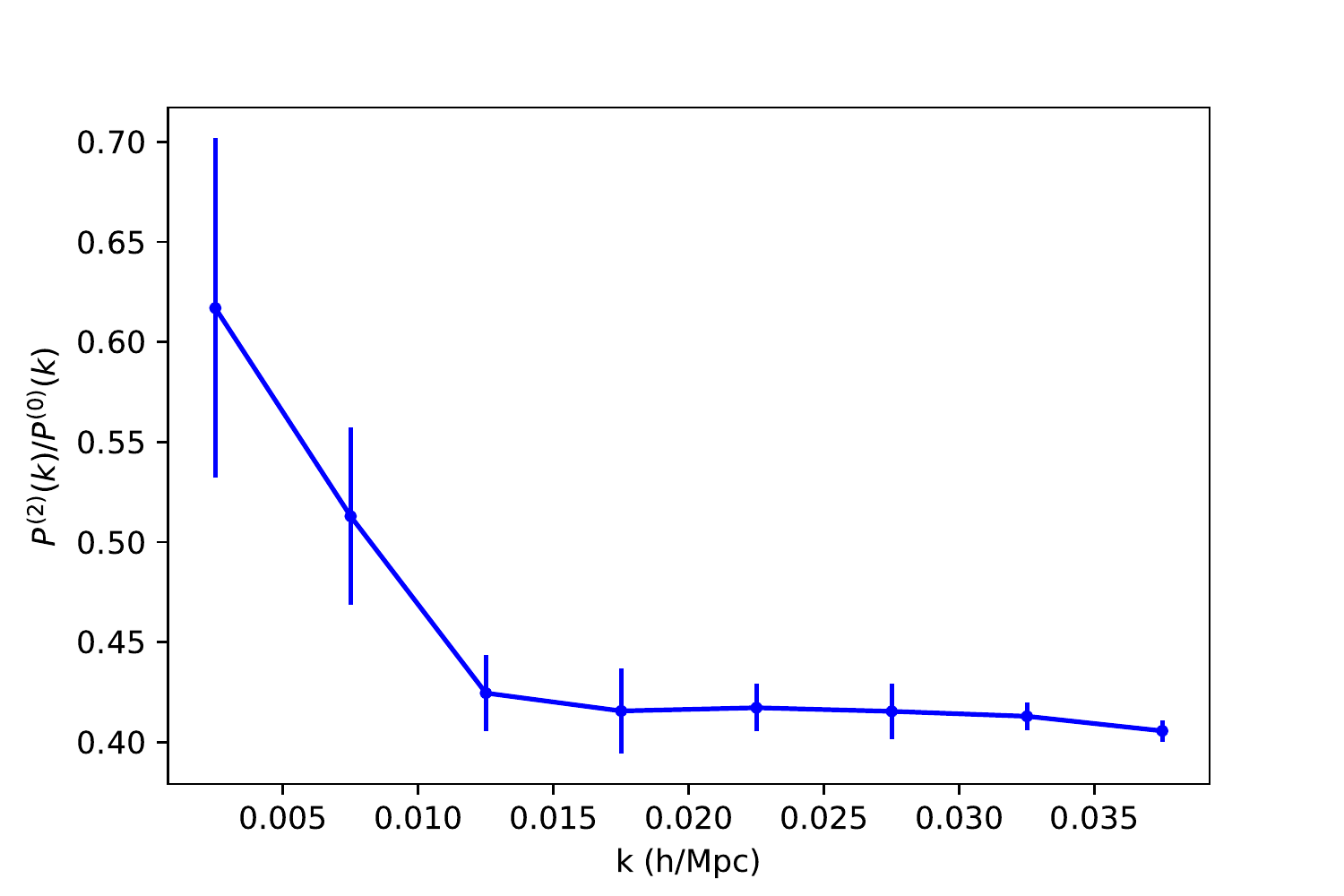}}
\caption{PS quadrupole to monopole ratio for halos of $M_{\rm min} =10^{13}h^{-1}\, M_{\odot}$  at $z=0$.}
\label{fig:kais}
\end{figure}

More effective constraints on $\beta_\alpha$ can be obtained by combining  the CR's with measurements of the ratio between the PS quadrupole and monopole, in the linear regime, see the second of  Eqs.~\re{PSmonquad}.
 This relation is valid in the Kaiser approximation \cite{Kaiser:1987qv}, whose validity is limited to small $k's$. Therefore we will fit this ratio only up to $k_{max}^{K}\alt 0.02-0.03$ $\hmp$, where the ratio exhibits the plateau shown in Fig.~\ref{fig:kais}. Notice that in deriving the CR's we have assumed the validity of linear theory, and therefore of the Kaiser approximation, up to $\qm$, so this procedure will be consistent as long as  $k_{max}^{K} \alt \qm$. Therefore, we will add to \re{chiCR} the function
\beq
\chi^2_K(\beta_\alpha) = \sum_j \frac{r_K^2(\beta_\alpha;k_j)}{\sigma_{K,j}^2}\,,
\eeq
where
\beq
r_K(\beta_\alpha;k_j) =\frac{P_\alpha^{(2)}(k_j)}{P_\alpha^{(0)}(k_j)} -   \frac{4 \beta_\alpha }{21} \frac{7+3\beta_\alpha}{1+\frac{2}{3} \beta_\alpha +\frac{1}{5}\beta_\alpha^2}\,,
\label{rk}
\eeq
and $\sigma_{K,j}$ is the error on the ratio between the PS quadrupole and monopole.  We will show combined constraints obtained by minimizing the sum
\beq
\chi^2_{TOT}=\chi^2_{CR}(b_\alpha,\beta_\alpha,\{a^{(l)}_i(q)\},\qm)+\chi^2_K(\beta_\alpha)\,.
\label{chitot}
\eeq
Summarizing, in our analysis we have 9 parameters, the physical ones \{$b_\alpha, \beta_\alpha$\} and the fitting ones \{$a_{-2}^{(l)},a_{0}^{(l)},a_{2}^{(l)},c$\}, with $l=1, 2$, over which we will marginalize.

\begin{figure}[htbp]
\center\includegraphics[width=0.4\textwidth]{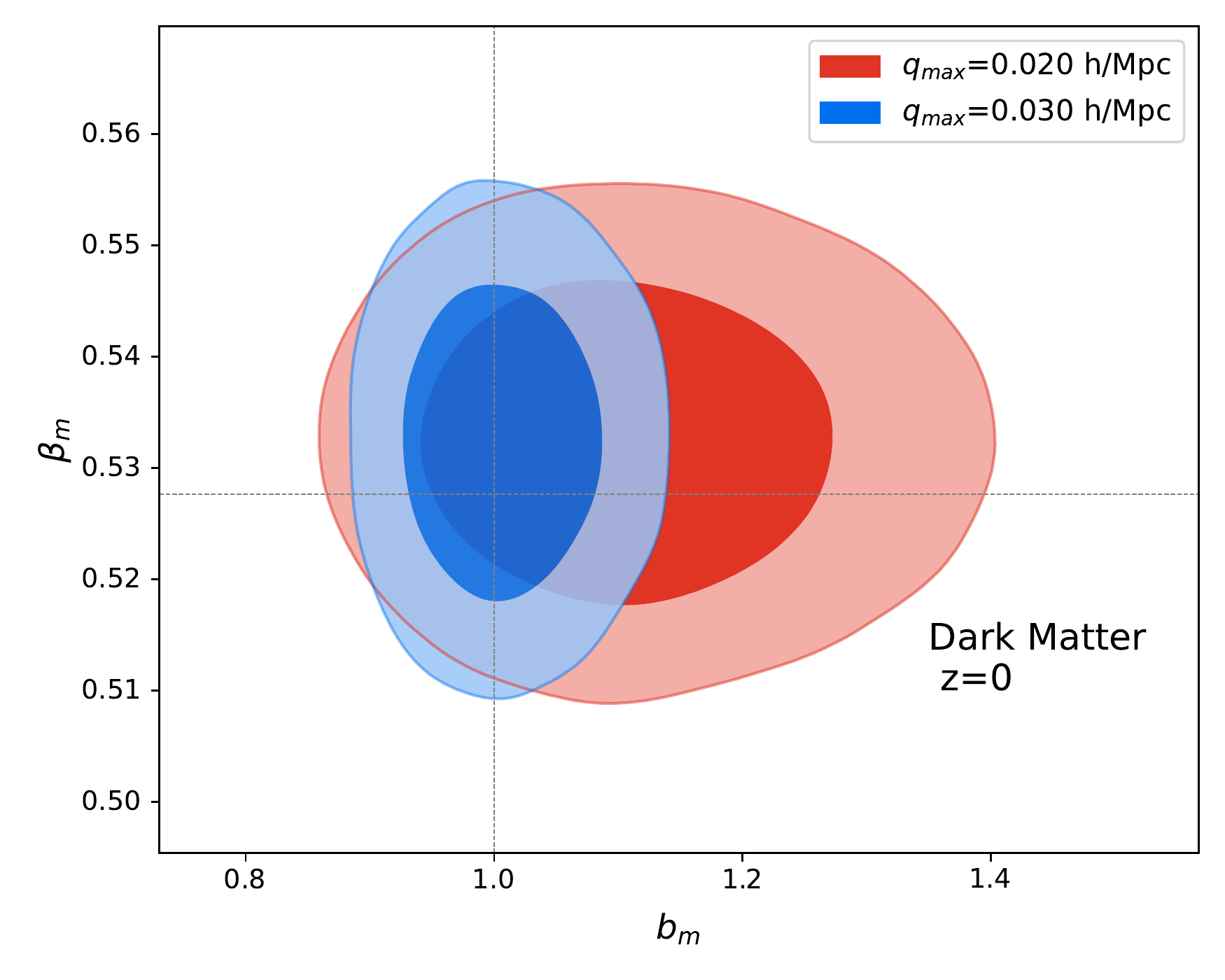}
\includegraphics[width=0.4\textwidth]{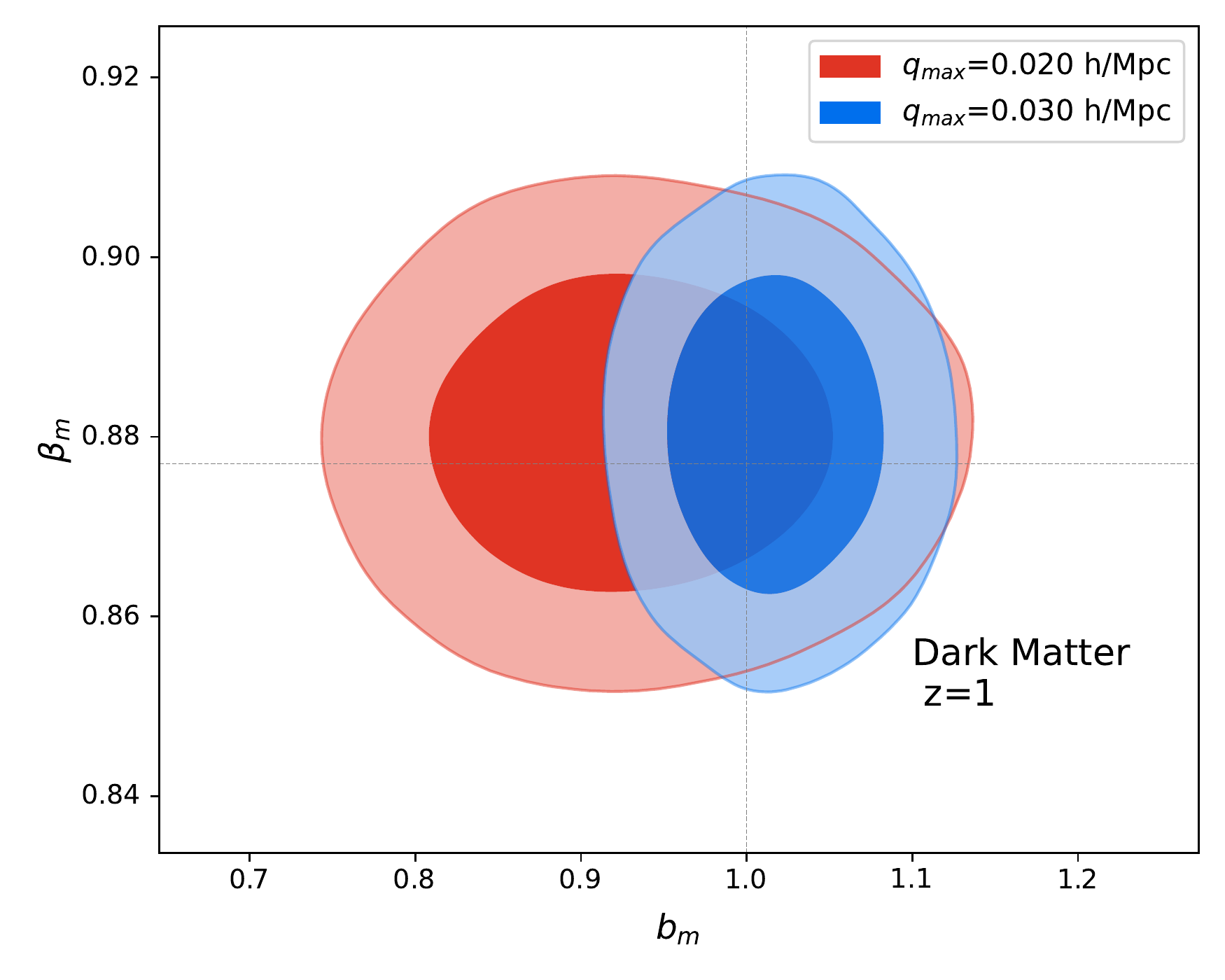}
\caption{1- and 2-$\sigma$ constraints in the $b_m - \beta_m$ plane for dark matter at redshift $z=0$ (left) and $z=1$ (right),  and for  different values of $q_{max}$. The dotted lines are the expected values for $b_m$ and $\beta_m$. }
\label{fig:matt_q}
\end{figure}

We first check our procedure for matter, for which we expect to extract values compatible with the fiducial ones,  $b_m^{fid}=1$ and $\beta_m^{fid}=f^{fid}$. We sample the log-likelihood function \re{chitot}  using the MCMC Python library \texttt{emcee}\footnote{https://emcee.readthedocs.io/en/stable/}\cite{ForemanMackey:2012ig}. The results of the analysis for matter are shown in Table~\ref{tab:DM_measurments} and in Fig.~\ref{fig:matt_q}, the plots are obtained using the plot library of Getdist\footnote{https://getdist.readthedocs.io/en/latest/intro.html}\cite{Lewis:2019xzd}.

As the constraining power of CR comes from the BAO's, we choose $k$ values in the range in which they are present in the bispectra. For dark matter we take $k_{min}=0.045$ $\hmp$ and $k_{max}=0.30$ $\hmp$. Higher values of $k_{max}$ do not improve our determinations of $b_\alpha$ and $\beta_\alpha$.
\\
In Fig.~\ref{fig:matt_q} we show the $68\,\%$ and $95\,\%$ confidence level regions in the $b_m - \beta_m$ plane for two different values of $\qm$ ($=0.02, \, 0.03\, \hmp$), and, with dotted lines, the fiducial values.   As we see, increasing $\qm$ improves the constraints, due to the higher number of triangular configurations included in the BS measurement. Both values of $\qm$ give unbiased values for the parameters. This is not the case by taking $\qm=0.04 \,\hmp$, which shows that this value is too far from the squeezed limit, as could have been anticipated also by looking at Figs.~\ref{residuals} and \ref{fits}.  Therefore, in our analysis on halos we will consider only $\qm=0.02,\, 0.03\;\hmp$.

\begin{table}[h]
\center \begin{tabular}{| c | c | c | c | c |}
\hline
\multicolumn{5}{| c |}{$z=0$ }\\
\hline
$q_{max}$ (h/Mpc) & $b_m$ & $b_m^{fid}$ & $f=\beta_m$ & $f^{fid}$\\ [0.5ex]
\hline\hline
0.020 & $1.11^{+0.12}_{-0.10}$ & 1 & $0.532^{+0.010}_{-0.009}$ & 0.528\\
\hline
0.030 & $1.01^{+0.05}_{-0.05}$ & 1 & $0.533^{+0.009}_{-0.009}$ & 0.528\\
\hline
\hline
\hline
\multicolumn{5}{| c |}{$z=1$}\\
\hline
$q_{max}$ (h/Mpc) & $b_m$ & $b_m^{fid}$ & $f=\beta_m$ & $f^{fid}$\\ [0.5ex]
\hline\hline
0.020 & $0.99^{+0.09}_{-0.08}$ & 1 & $0.880^{+0.012}_{-0.012}$ & 0.877\\
\hline
0.030 & $1.02^{+0.04}_{-0.04}$ & 1 & $0.880^{+0.012}_{-0.012}$ & 0.877\\
\hline
\end{tabular}
\caption{Best fit values for $b_m$ and $f$ at different redshifts for different values of the maximum allowed $q$. For each parameter,  the quoted  errors correspond to  the 68\% CL. of the one-dimensional probability distribution function.}
\label{tab:DM_measurments}
\end{table}

\begin{center}
\begin{figure}[htbp]
\includegraphics[width=0.33\textwidth]{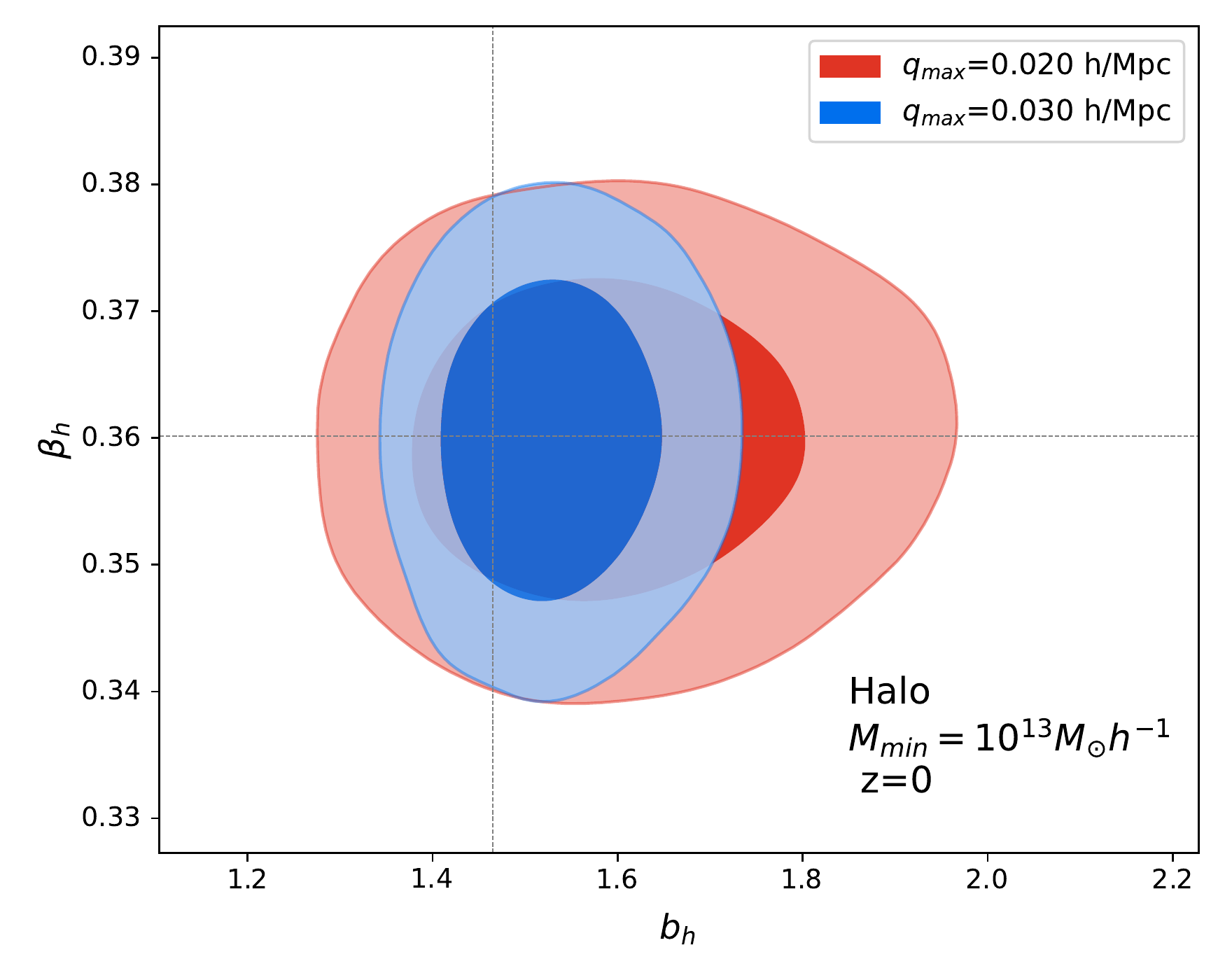}
\includegraphics[width=0.33\textwidth]{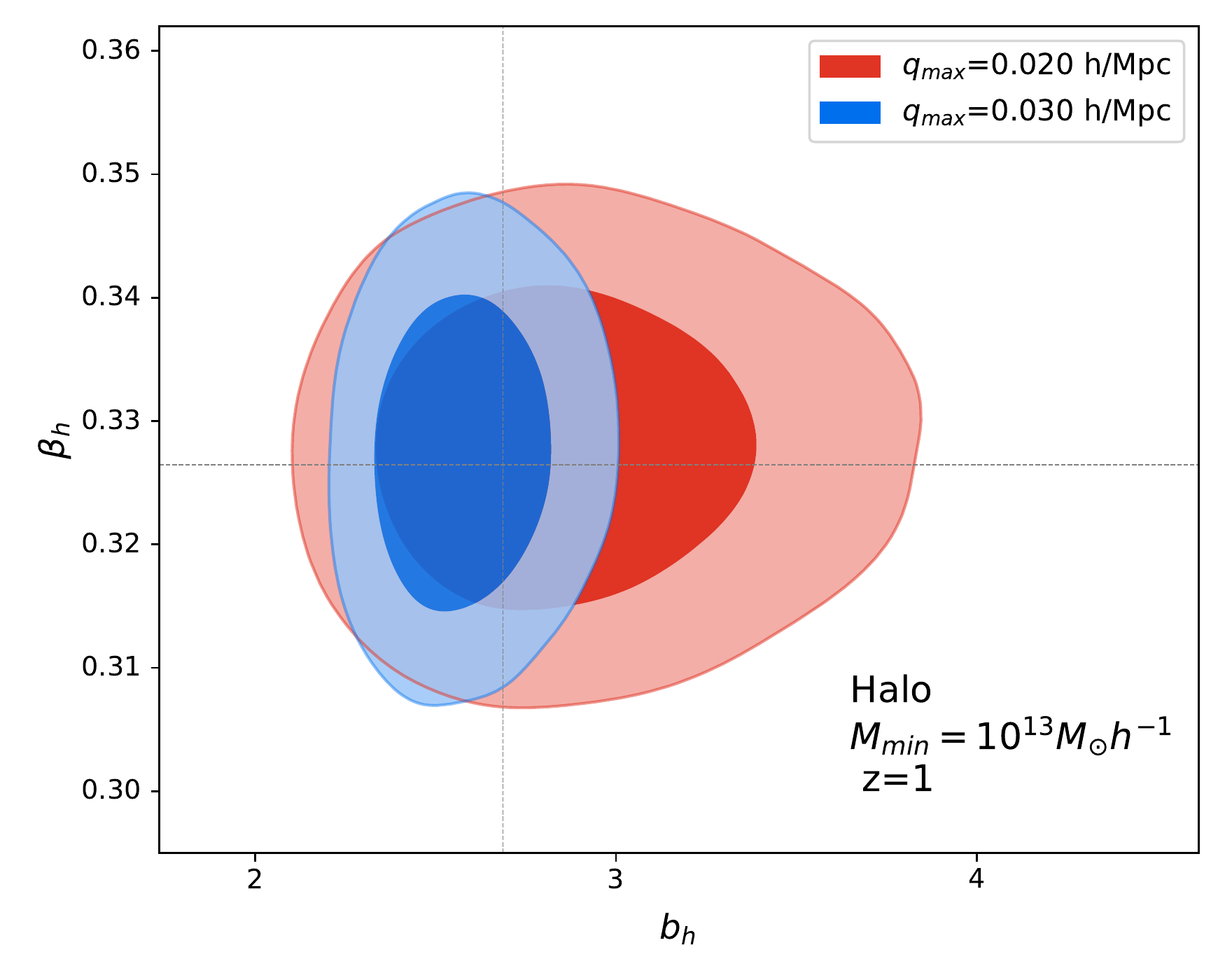}
\includegraphics[width=0.33\textwidth]{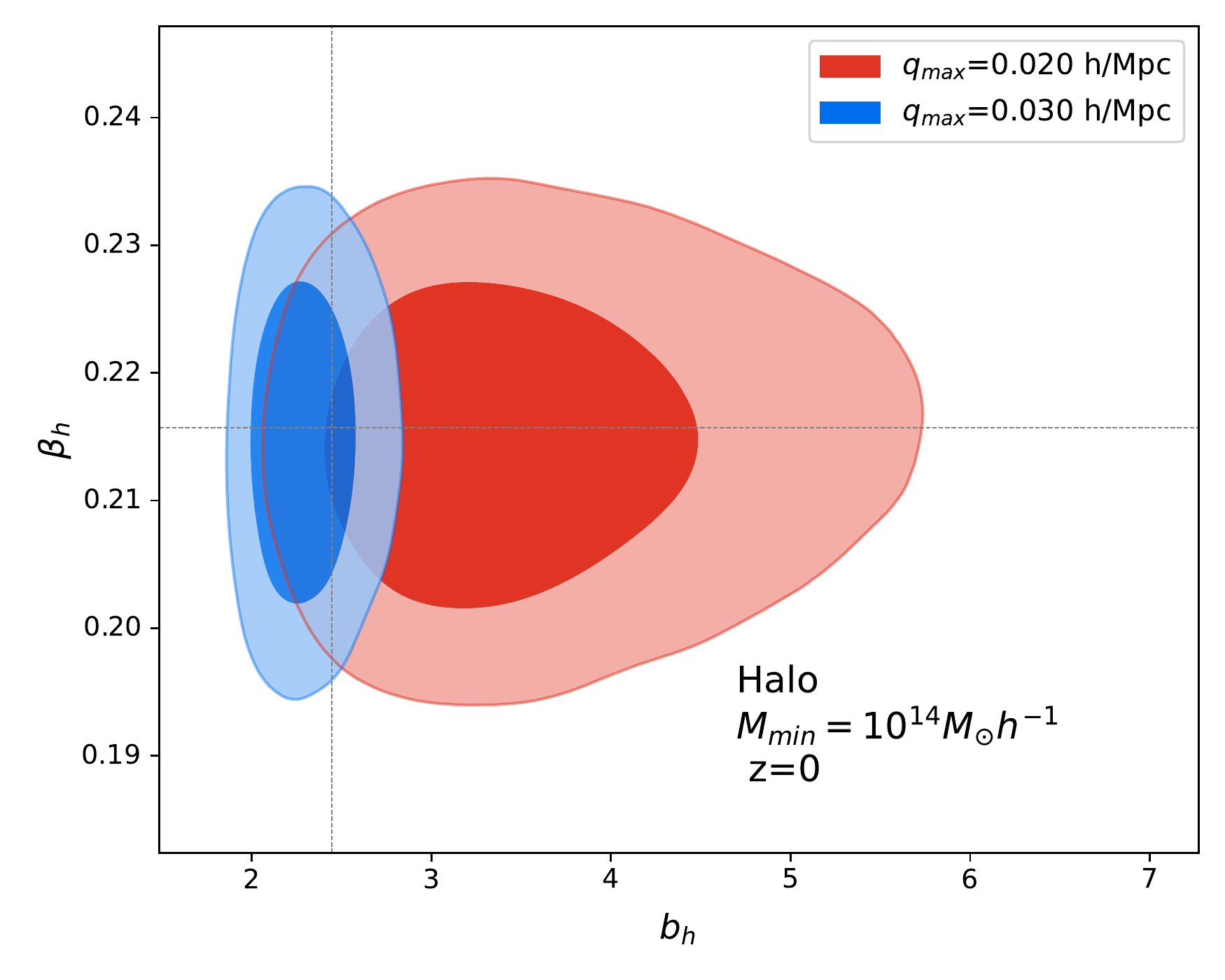}
\caption{1- and 2-$\sigma$ constraints on $b_h$ and $\beta_h$ for halos of $M_{\rm min} =10^{13}h^{-1}\, M_{\odot}$  at $z=0$ (right) and $z=1$ (middle)   $M_{\rm min} =10^{14} h^{-1}\,M_{\odot}$  at $z=0$ (right)  for  two different values of $q_{max}$ ($=0.02,\,0.03\,\hmp$). The dotted lines are the expected values, obtained from direct measurements. }
\label{fig:halo_q}
\end{figure}
\end{center}

After having validated the procedure for dark matter we proceed in the analysis for halos with different masses at different redshifts. The results are presented in Table~\ref{tab:halo_measurments} and Fig.~\ref{fig:halo_q} and are evaluated using $k_{min}(z=0) = 0.035$~$\hmp$, $k_{max}(z=0)=0.26$~$\hmp$ and $k_{min}(z=1)=0.035$~$\hmp$, $k_{max}(z=1)=0.28$~$\hmp$. The results are  compatible with the theoretical fiducial values at the $1-\sigma$ level, with the expected values measured from the simulations   using Eq.~\re{bias} for $b_h^{fid}$, and Eq.~\re{PSmonquad} for $\beta_h^{fid}$.

\begin{table}[h]
\center
\begin{tabular}{| c | c | c | c | c |}
\hline
\multicolumn{5}{| c |}{ $M_{\rm min} =10^{13} h^{-1}\,M_{\odot} \qquad z=0$ }\\
\hline
$q_{max}$ (h/Mpc) & $b_h$ & $b_h^{fid}$ & $f=\beta_h b_h$ & $f^{fid}$\\ [0.5ex]
\hline\hline
0.020 & $1.58^{+0.15}_{-0.13}$ & 1.47 & $0.57^{+0.06}_{-0.06}$ & 0.528\\
\hline
0.030 & $1.53^{+0.08}_{-0.08}$ & 1.47 & $0.55^{+0.04}_{-0.04}$ & 0.538\\
\hline
\hline
\hline
\multicolumn{5}{| c |}{$M_{\rm min} =10^{13} h^{-1}\,M_{\odot} \qquad z=1$ }\\
\hline
$q_{max}$ (h/Mpc) & $b_h$ & $b_h^{fid}$ & $f=\beta_h b_h$ & $f^{fid}$\\ [0.5ex]
\hline\hline
0.020 & $2.85^{+0.39}_{-0.32}$ & 2.686 & $0.93^{+0.14}_{-0.14}$ & 0.877\\
\hline
0.030 & $2.58^{+0.17}_{-0.16}$ & 2.686 & $0.82^{+0.07}_{-0.07}$ & 0.877\\
\hline
\hline
\hline
\multicolumn{5}{| c |}{$M_{\rm min} =10^{14} h^{-1}\,M_{\odot} \qquad z=0$ }\\
\hline
$q_{max}$ (h/Mpc) & $b_h$ & $b_h^{fid}$ & $f=\beta_h b_h$ & $f^{fid}$\\ [0.5ex]
\hline\hline
0.020 & $3.40^{+0.83}_{-0.61}$ & 2.446 & $0.73^{+0.18}_{-0.18}$ & 0.528\\
\hline
0.030 & $2.29^{+0.21}_{-0.18}$ & 2.446 & $0.49^{+0.06}_{-0.06}$ & 0.528\\
\hline
\end{tabular}
\caption{Determination of $b_h$ and $\beta_h$ from the CR's for halos, for different values of $\qm$. For each parameter,  the quoted  errors correspond to  the 68\% CL. of the one-dimensional probability distribution function.}
\label{tab:halo_measurments}
\end{table}

Our analysis shows that it is possible to break the degeneracy between the linear bias and the growth rate (or the $\beta$-parameter) with a good accuracy. In Table~\ref{tab:DM_measurments}  and Table~\ref{tab:halo_measurments} we report the 68~\% CL measurements we obtained for $q_{max}=0.02,0.03$ $\hmp$. We can see that the results of the analysis  for biased tracers in redshift space are fully consistent with the fiducial $\Lambda$CDM  value for the growth function $f$.

Notice that, both for matter and halos, CR constrain mainly the bias parameter $b_\alpha$, while the parameter $\beta_\alpha$ is constrained mostly by the independent measurement of the PS quadrupole to monopole ratio. As the latter is measured  at better than $5\,\%$ accuracy for our simulations, the error on the derived growth function $f$ is dominated by that on the bias parameter.

\section{Estimating constraining power}
\label{forecast}
We here present a forecast of the expected constraining power using the CR on the oscillatory part of the spectra alone, that is, not in conjunction with the PS quadrupole to monopole ratio. Since we have seen that the impact of the redshift space distortion is rather weak, a separate constraint on the tracer bias $b_\alpha$ and the growth rate parameter $f$ from the consistency relation alone would be difficult. We thus focus on the constraint on $b_\alpha$ ignoring the redshift space distortions, that is, setting $f=0$ in the CR's. As discussed earlier, we can then combine with the constraint on $\beta_\alpha$ from the redshift space distortion on the PS, to disentangle the degeneracy between the two parameters. Another simplification that we have made for the forecast is to ignore nonlinear damping of BAOs. Including this effect would weaken the constraint especially from high wavenumbers, and thus the results presented here would give us the best-case scenario, but the purpose here is to give a rough idea on the statistical power brought by the consistency relations and the simplified treatment here must be fine.

We start with the construction of the BAO template based on the linear matter PS. As discussed in Sect.~\ref{resid}, we take the logarithmic derivative, $d \ln P^0(k)/d\ln k$, and then subtract a B-spline fit to extract the oscillatory part. We use this as the template model after multiplying by $(1/3b_\mathrm{g}) P_\mathrm{g}(q)P_\mathrm{g}(k)$ for the monopole and by $(2/3b_\mathrm{g}) P_\mathrm{g}(q)P_\mathrm{g}(k)$ for the quadrupole of the galaxy BS, where $P_\mathrm{g}(k)=b_\mathrm{g}^2P^0(k)$ is the linear galaxy PS with the bias parameter $b_\mathrm{g}$. We then estimate the covariance matrix of the galaxy BS, which is diagonal under the Gaussian assumption \cite{Scoccimarro_1998}:
\begin{eqnarray}
\left[\Delta B_\mathrm{g}(k_1,k_2,k_3)\right]^2 = \frac{V}{N_\mathrm{tri}}\left[P_\mathrm{g}(k_1)+n_\mathrm{g}^{-1}\right]\left[P_\mathrm{g}(k_2)+n_\mathrm{g}^{-1}\right]\left[P_\mathrm{g}(k_3)+n_\mathrm{g}^{-1}\right],\nonumber\\
\label{eq:error_bispec}
\end{eqnarray}
where $N_\mathrm{tri}$ is the number of Fourier triangles in a bin, which scales as $V^2$, and $n_\mathrm{g}$ is the galaxy number density specified later assuming a future survey setting. Since we specify the triangles by $(q,k)$ and average over the angular dependence in our case, $N_\mathrm{tri}$ after taking this average can be expressed as
\begin{eqnarray}
N_\mathrm{tri} = \left(\frac{2\pi}{3}\right)^2\frac{(q_\mathrm{bin,max}^3-q_\mathrm{bin,min}^3)(k_\mathrm{bin,max}^3-k_\mathrm{bin,min}^3)}{k_\mathrm{f}^6},
\end{eqnarray}
where $q_\mathrm{bin,min}$ and $q_\mathrm{bin,max}$ specify the minimum and the maximum wavenumber of the $q$ bin and similarly for the $k$ bin. We adopt the bin spacing of $0.005\,\hmp$ for this forecast, and have confirmed that the results are virtually unchanged when we adopt a finner binning. In the above, we have excluded the contribution from redundant triangles (e.g., a triangle with negative $q_z$ is equivalent to another with positive $q_z$) due to the reality condition, $\delta_{-{\mathbf{q}}} = \delta_{\mathbf{q}}^*$, and denote the fundametal wavenumber by $k_\mathrm{f}=2\pi/V^{1/3}$.
The error on the monopole moment of the BS is estimated using Eq.~\re{eq:error_bispec} assuming that $P^0(k_1)=P^0(q)$ and $P^0(k_2)=P^0(k_3)=P^0(k)$ approximately hold over the triangles in a bin, and that of the quadrupole is obtained by further multiplying a factor $5$ to Eq.~\re{eq:error_bispec} to account for the weighting by the Legendre polynomial and our normalization convention. We ignore the error on the PS, which should be much smaller than that in the BS.

We consider a Euclid-like survey and take the survey parameters from Table 3 in Ref.~\cite{Amendola:2016saw}. Instead of considering the tomographic analysis with the $14$ thin redshift bins over $0.65<z<2.05$ listed in that table, we consider three thick redshift bins with similar volume as summarized in Table~\ref{tab:param_forecast}. We consider the survey area of $15,000\,\mathrm{deg}^2$ and take the values in ``reference'' case for the galaxy number density, averaged over the relevant fine redshift bins weighted by the volume. We propagate the error on the monopole and the quadrupole moment of the bispectra to the only parameter of the model template, $b_\mathrm{g}$, to give the estimate of the statistical power of the consistency relation. We fix $b_\mathrm{g}=1.5$ as the fiducial value for all the three bins. Changing this would give us a slight change in the relative contribution of the shot noise, but the final forecast is almost unchanged when we modify this to e.g., $b_\mathrm{g}=1.6$.

\begin{table}[t!]
    \caption{Survey parameters considered in the forecast. \label{tab:param_forecast}}
    \begin{center}
  \begin{tabular}{c c c}
    redshift & $V\,[h^{-3}\mathrm{Gpc}^3]$ & $(n_\mathrm{g}/10^{-4})\,[h^{3}\mathrm{Mpc}^{-3}]$\\
    \hline
    $0.65<z<1.25$ & 22.64 & 15.86\\
    $1.25<z<1.65$ & 20.66 & 8.86\\
    $1.65<z<2.05$ & 22.69 & 2.61
  \end{tabular}
\end{center}
\end{table}

\begin{center}
\begin{figure}[htbp]
\includegraphics[width=0.33\textwidth]{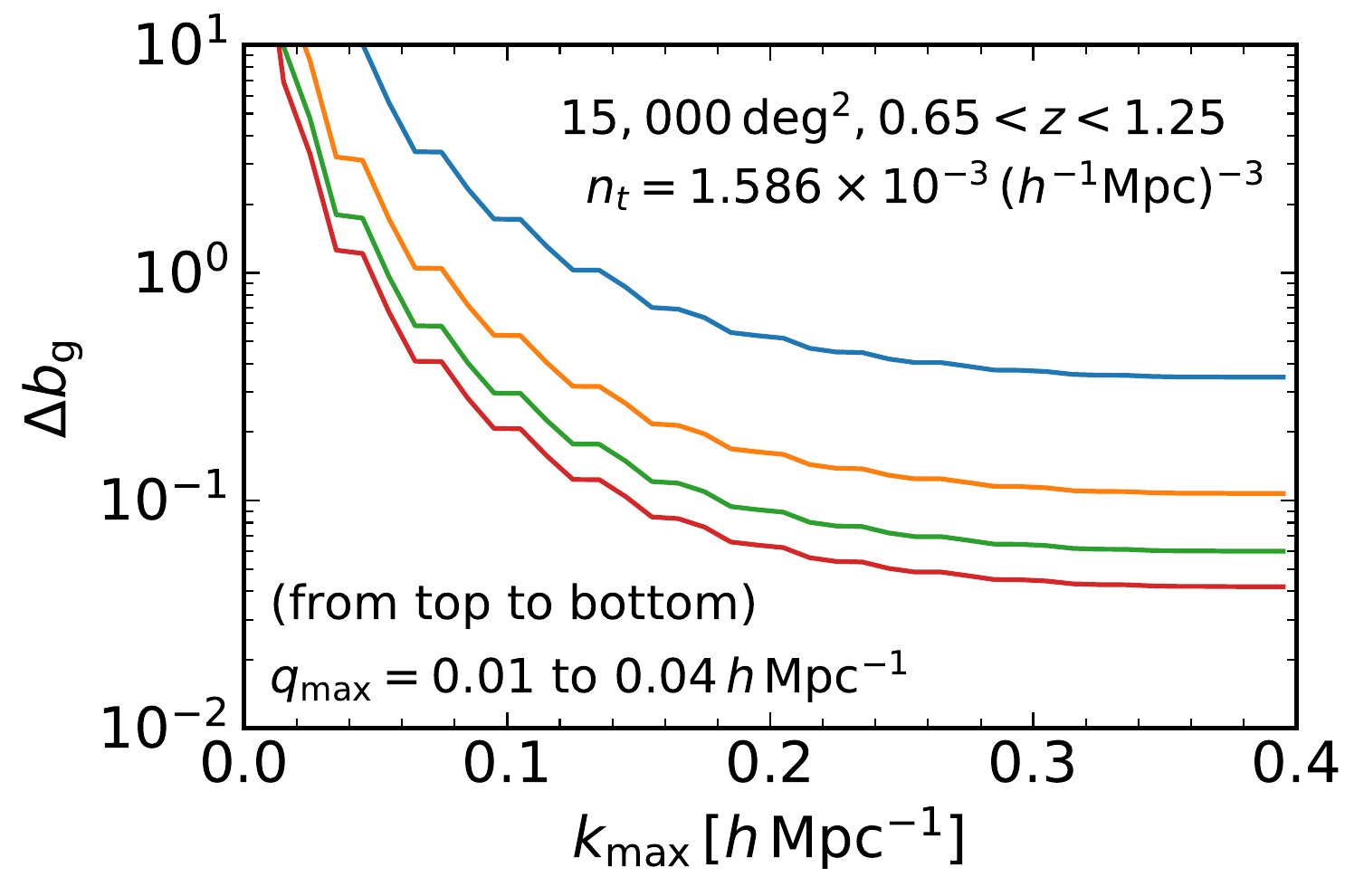}
\includegraphics[width=0.33\textwidth]{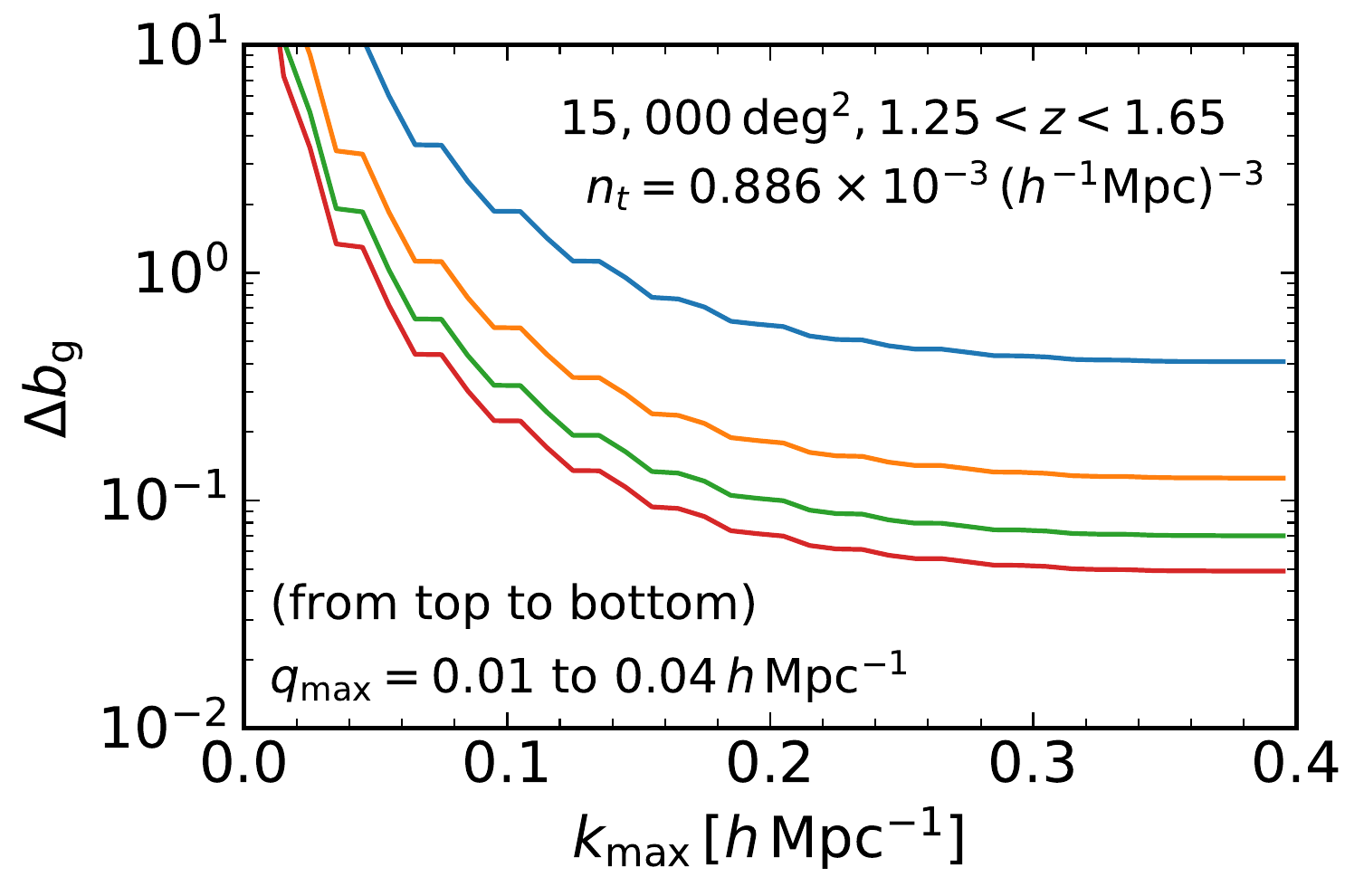}
\includegraphics[width=0.33\textwidth]{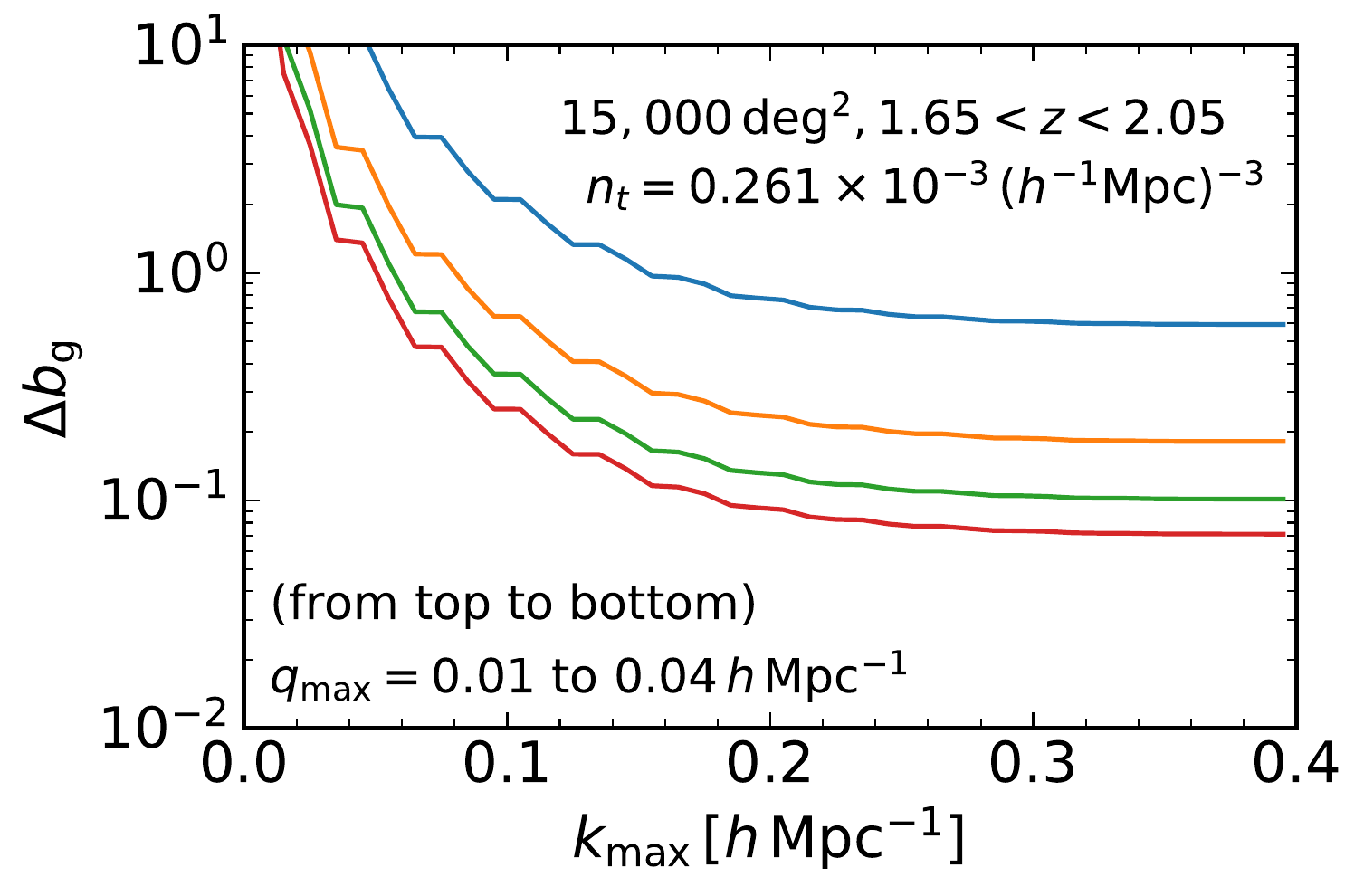}
\caption{Forecast of the constraint on the galaxy bias parameter $b_\mathrm{g}$ from a Euclid-like survey in three tomographic redshift bins coming from the oscillatory part of the consistency relation alone. The results are shown as a function of the maximum hard wavenumber $k_\mathrm{max}$ for some values of the corresponding soft wavenumber limit, $q_\mathrm{max}$.}
\label{fig:forecast}
\end{figure}
\end{center}

We show in Fig.~\ref{fig:forecast} the expected $1$-$\sigma$ error on the bias parameter as a function of the maximum wavenumber included in the analysis. While the limit of the hard wavenumber, $k_\mathrm{max}$, is indicated 
by the $x$-axis, we consider four values of $q_\mathrm{max}$, the counterpart for the soft wavenumber, $0.01, 0.02, 0.03$ and $0.04\,\hmp$. The smaller $q_\mathrm{max}$ is, we are restricting to more squeezed triangles and the resultant constraint is weaker. As we already see explicitly in previous sections, we can push to $q_\mathrm{max}$ to $0.03\,\hmp$ quite safely without introducing a sizable bias in the consistency relation. While $q_\mathrm{max}=0.04\,\hmp$ might be slightly optimistic the improvement from $q_\mathrm{max}=0.03\,\hmp$ is smaller compared to that from $q_\mathrm{max}=0.02\,\hmp$ to $q_\mathrm{max}=0.03\,\hmp$.

It is clear from the figure that we can achieve a better than ten percent determination of the bias parameter for all the redshift bins, with the highest redshift bin slightly worse due to the larger shot noise error. Since the nonlinear damping of BAO is not very significant for scales $k<0.1\,\hmp$ and the most of the constraining power is coming from the $k<0.2\,\hmp$, above which the shot noise error gets prominent, our estimate should be a good approximation even when the nonlinear effects are considered.

\section{Conclusions}
\label{concl}
In this paper, we have investigated the CR's  as a way to measure the large scale bias and the large scale growth rate in a model independent way. We have derived the relevant CR's in redshift space for the BS monopole and quadrupole and verified their validity on a set of large volume 
N-body simulations, both for DM and for haloes of different mass, at different redshifts.
While the coefficients of the CR's depend on $b_\alpha$ and $\beta_\alpha$ separately, the constraining power on  $\beta_\alpha$  from CR's alone turns out to be very mild. However, when the CR measurements are combined with those on the PS quadrupole to monopole ratio, the $b_\alpha-\beta_\alpha$ degeneracy is completely broken.

When applied to a Euclid-like survey this approach would provide constraints on these parameters at 
better than $10\%$ level. It is likely that this result can be further improved by modelling the leading contributions not protected by the CR's, and we think it will be very interesting to explore quantitatively this issue.
In any case, while a ten percent error would not sound to be particularly good in modern cosmology, our constraints come completely free from model assumptions given as a bonus by just checking certain configurations of the BS. This, when combined with the redshift space distortion, which cannot break the degeneracy between the bias and the growth-rate parameter, would provide a unique way to constrain the gravitational growth.

Considering different redshift bins, the extracted values for $f(z)$ would help constraining $\Lambda$CDM and modified scenarios as well. Having multiple  tracers
available would provide a unique way of testing the universality of the large scale growth rate, constraining possible velocity bias and violations of the EP.  We leave the exploration of these applications to future work.

\section*{Acknowledgments}
MP acknowledges support from the European Union Horizon 2020 research and innovation programme under the Marie Sklodowska-Curie grant agreements Invisible- sPlus RISE No. 690575, Elusives ITN No. 674896 and Invisibles ITN No. 289442. TN was supported in part by World Premier International Research Center Initiative (WPI Initiative), MEXT, Japan, JSPS KAKENHI Grant Numbers JP17K14273 and JP19H00677, and by JST AIP Acceleration Research Grant Number JP20317829, Japan.
The numerical simulations and subsequent postprocessing were carried out on Cray XC50 at Center for Computational Astrophysics, National Astronomical Observatory of Japan.

\appendix
\section{General Derivation of the CR's}
\label{gender}

In this section, we derive the CR's in a way convenient for the purpose of this paper, and, moreover, we specify to the equal-time limit, which was not treated in the original papers \cite{Peloso:2013zw, Kehagias:2013yd}.

We consider the most general Boltzmann equation,
\beq
\left(\frac{\partial}{\partial \tau}  + \frac{p_i}{a m} \frac{\partial }{\partial x^i} - a m \frac{\partial}{\partial x^i}\Phi(\bx,\tau) \frac{\partial}{\partial p_i } \right) f(\bx, \bp, \tau) ={\cal C}[f,\ldots](\bx, \bp, \tau)\,,
\label{Boltzmann}
\eeq
where $f(\bx, \bp, \tau)$ is the distribution function of a given species, not necessarily cold dark matter. The collision term at the RHS, takes into account possible non-gravitational interactions, and it involves $f$ itself as well as the distribution functions of the other species taking part in the interactions. $f$ could also represent the distribution of halos in a given mass range, or a given type of galaxies, and in that case ${\cal C}$ would describe processes which change the comoving number density of these tracers, such as merging. Eq.~\re{Boltzmann} is invariant under the
time-dependent frame change
\beq
\bx\to \bar \bx = \bx+ \bd(\tau)\,,\qquad \bp \to\bar\bp=\bp+a m \dot \bd(\tau)\,,
\label{shift}
\eeq
provided we make the replacements
\beqra
&&f(\bx, \bp, \tau)\to \bar f(\bx, \bp, \tau)= f(\bx-\bd(\tau),\bp-a m \dot \bd(\tau),\tau )\,,\nonumber\\
&& \frac{\partial}{\partial x^i}\Phi(\bx,\tau)\to\frac{\partial}{\partial x^i}\bar\Phi(\bx,\tau) =\frac{\partial}{\partial x^i}\Phi(\bx-\bd(\tau),\tau) -{\cal H} \dot\bd(\tau) -\ddot\bd(\tau)\,,
\label{shift2}
\eeqra
and if the collisional term satisfies
\beq
{\cal C}[\bar f,\ldots]( \bx,  \bp, \tau)={\cal C}[ f,\ldots](\bx-\bd(\tau),  \bp-a m \dot \bd(\tau), \tau) \,,
\label{shift3}
\eeq
that is, the interaction rate is the same in the two frames.
The transformation above is nothing but the Equivalence Principle (EP), also called in this context the extended galilean invariance.
Since it is  an invariance of the Boltzmann equation, its consequences are not restricted to perturbation theory, but are valid at the fully nonlinear level, also including nonperturbative effects such as shell-crossing and multistreaming. Moreover, one should keep in mind that the symmetry holds for an {\em arbitrary} displacement $\bd(\tau)$, independently on the identification of it as the infinite wavelength limit of a large scale cosmological perturbation. The last observation is crucial in order to disentangle the {\em dynamical} content of the consistency relation from the {\em statistical} one, related to the statistical properties of the cosmological perturbations such as adiabaticity and gaussianity.

The dynamical content is encoded in constraints on the mode-coupling vertices in the {\em soft} limit, that is, when one of the modes goes to zero. It is best analyzed in Fourier space, by replacing the homogeneous displacement $\bd(\tau)$ with a scale dependent one,
\beq
\int d^3 x\,e^{i\bx\cdot\bq} \bd(\tau) = (2\pi)^3 \delta_D(\bq)  \bd(\tau) \to \tilde\bd(\bq,\tau)\,,
\label{soft}
\eeq
and then considering the $q\to 0$ limit.

We will focus on the BS in redshift space
\beqra
 &&\!\!\!\!\!\!\!\!\!\!\! B^{(S)}_{\alpha\beta\gamma}(\bq,-\bk_+,\bk_-;\tau_\alpha,\tau_\beta,\tau_\gamma) \equiv \langle \delta^{(S)}_\alpha(\bq;\tau_\alpha) \delta^{(S)}_\beta(-\bk_+;\tau_\beta) \delta^{(S)}_\gamma(\bk_-;\tau_\gamma)  \rangle^\prime\,,
 \label{bisp}
 \eeqra
  where $\bk_\pm=\bk\pm\frac{\bq}{2}$, $q=|\bq|$, $k_\pm=|\bk_\pm|$,  and the prime indicates that the expectation value has been divided by a $(2\pi)^3 \delta_D(0)$ factor. $\delta^{(S)}_{\alpha, \beta, \gamma}$ indicate the density contrasts for different tracers (e.g. DM, baryons, a given galaxy type, ...), evaluated at times $\tau_{\alpha,\beta,\gamma}$, respectively.

By moving to another frame, the transformations \re{shift}-\re{shift3} dictate the transformation of the BS. The density contrasts (obtained from the first moments of the distribution function) transform as
\beq
\!\!\!\!\!\!\!\!\!\!\!\!\! \!\!\!\!\!\!\!\!\!  \delta^{(S)}(\bk ,\tau)\to \bar\delta^{(S)}(\bk,\tau) = \delta^{(S)}(\bk,\tau) +\, i I_{\bk;\bq^\prime,\bp} \bp\cdot  \tilde \bd(\bq^\prime,\tau)  \delta^{(S)}(\bp,\tau) +\cdots,
\eeq
where
\beq
I_{\bk;\bp_1,\bp_2} \equiv \int\frac{d^3 p_1}{(2\pi)^3}\frac{d^3 p_2}{(2\pi)^3} (2\pi)\delta_D(\bk-\bp_1-\bp_2)\,,
\eeq
and the dots indicate higher orders in $ \tilde \bd$.
Inserting it in \re{bisp} the additional  contributions to the BS induced by the change of frame are obtained,
\beqra
&& i I_{-\bk_+;\bq',\bp} \langle\bp\cdot \tilde\bd(\bq^\prime,\tau_\beta) \delta^{(S)}_\alpha(\bq,\tau_\alpha) \delta^{(S)}_\beta(\bp,\tau_\beta) \delta^{(S)}_\gamma(\bk_-;\tau_\gamma)   \rangle^\prime\nonumber\\
&&+(-\bk_+\leftrightarrow \bk_-\,,\;\;\;\; \beta \leftrightarrow \gamma) +(-\bk_+\leftrightarrow \bq,,\;\;\;\; \beta \leftrightarrow \alpha)\,,
\label{corr1}
\eeqra
where the two parentheses at the second line stand for two contributions obtained from the one at the first line by performing the replacements indicated.
When the uniform limit for the displacement (that is, the inverse of \re{soft}) is taken, the sum of the three new contributions gives the BS  itself multiplied by the coefficient
\beq
- i \left(\bk_+ \cdot \bd(\tau_\beta) - \bk_-\cdot \bd(\tau_\gamma) -\bq\cdot\bd(\tau_\alpha) \right)\,,
\label{etimes}
\eeq
which vanishes for $\tau_\alpha=\tau_\beta=\tau_\gamma$, as a consequence of the EP and translational invariance. This holds indeed at every order in $\bd$, as it was shown  in \cite{Scoccimarro:1995if}.

In order to obtain the CR's, one has to identify the displacement $\tilde \bd(\bq,\tau)$ with the large scale displacements induced by the velocity perturbations, that is, one has to give a {\it statistical} content to it. Assuming that at large scales linear PT holds, we will then identify, in redshift space,
\beq
\!\!\!\!\! \tilde \bd(\bq,\tau) = \frac{1}{{\cal H}f} \left(\bv(\bq,\tau)+f \bv(\bq,\tau)\cdot \bf{\hat z}  \;\bf{\hat z}\right) = -i \frac{\delta_m(\bq,\tau)}{q^2}\left(\bq+ f \,\bq \cdot \bf{\hat z}  \;\bf{\hat z} \right),
\label{lt}
\eeq
with $\delta_m(\bq,\tau)$  the real space matter density field, which is related to the velocity field by the continuity equation. Implicitly, we have assumed that all the different species fall, al large scales, with the same velocity field, which follows from the assumption of adiabatic initial conditions and, again, the EP.
Inserting \re{lt}  in the first term in \re{corr1} we get, in the $q/k\to 0$ limit,
\beq
\simeq \frac{k}{q} \left( \mu + f \mu_q\mu_k \right) \langle \delta^{(S)}_\alpha(\bq,\tau_\alpha) \delta_m(-\bq,\tau_\beta)\rangle^\prime \langle\delta^{(S)}_\beta(-\bk_-,\tau_\beta)\delta^{(S)}_\gamma(\bk_-,\tau_\gamma) \rangle^\prime\,,
\eeq
where
\beq
\mu \equiv\frac{\bk\cdot\bq}{kq}\,,\qquad \mu_q\equiv \frac{\bq\cdot\hat{\bz}}{q}\,,\qquad  \mu_k\equiv \frac{\bk\cdot\hat{\bz}}{k}\,.
\eeq
A contribution proportional to $k/q$ is obtained also from the second term in Eq.~\re{corr1}, while the third one vanishes. Consistently with our assumption that linear PT holds at the scale $q$, we use the Kaiser relation to express the real space matter field in terms of the redshift space one for the tracer $\alpha$,
\beq
\delta_m(\bq,\tau)= \frac{1}{b_\alpha + f \mu_q^2} \, \delta^{(S)}_\alpha(\bq,\tau)\,,
\eeq
where both $f$ and $b_\alpha$ are evaluated at the time $\tau$. So, combining with Eq.~\re{lt}, we have
\beq
\tilde \bd(\bq,\tau) =- \frac{i}{q^2} \frac{ \bq+ f \,\bq \cdot \bf{\hat z}  \;\bf{\hat z} }{b_\alpha + f \mu_q^2} \, \delta^{(S)}_\alpha(\bq,\tau)\,.
\label{lth}
\eeq
Finally, we get
 \beqra
 &&\!\!\!\!\!\!\!\!\!  \!\!\!\!\!\!\!\!\!  \!\!\!\!\!\!\!\!\! \lim_{q/k\to0} B^{(S)}_{\alpha\beta\gamma}(\bq,-\bk_+,\bk_-;\tau_\alpha,\tau_\beta,\tau_\gamma) \nonumber\\
&&\!\!\!\!\!\!\!\!\!  \!\!\!\!\!\!\!\!\!  \!\!\!\!\!\!\!\!\!  =-  \frac{k}{q}\frac{\mu + f \mu_k \mu_q}{b_\alpha + f\mu_q^2}   P^{(S)}_{\alpha \alpha}(\bq;\tau_\alpha,\tau_\alpha) \left[ \frac{D(\tau_\gamma)}{D(\tau_\alpha) }P^{(S)}_{\beta \gamma}(\bk_+;\tau_\beta,\tau_\gamma)-\frac{D(\tau_\beta)}{D(\tau_\alpha) }P^{(S)}_{\beta \gamma}(\bk_-;\tau_\beta,\tau_\gamma)
\right]\nonumber\\
&&\!\!\!\!\!\!\!\!\!  \!\!\!\!\!\!\!\!\!  \!\!\!\!\!\!\!\!\!+ O\left(\left(\frac{q}{k}\right)^0\right)\,,
\label{CR_diff_t}
 \eeqra
 where the PS's are defined as
 \beq
 P^{(S)}_{\alpha\beta}(\bk;\tau_\alpha,\tau_\beta)\equiv\langle \delta^{(S)}_\alpha(\bk;\tau_\alpha)\delta^{(S)}_\beta(-\bk;\tau_\beta) \rangle^\prime\,,
 \eeq
 $D(\tau)$ is the linear matter growth factor and we have assumed the linear behavior of the PS time dependence at the {\em soft} scale $q$,  $P^{(S)}_{\alpha \alpha}(\bq;\tau_\alpha,\tau_\beta)=P^{(S)}_{\alpha \alpha}(\bq;\tau_\alpha,\tau_\alpha) D(\tau_\beta)/D(\tau_\alpha)$. On the other hand,
as we have already  emphasized, the dynamics at the {\it hard} scale, $k$ is completely nonlinear.  The key point is that the structure of the first term at the RHS is protected against any kind of, perturbative and nonperturbative, nonlinear effect. By contrast, the form of the remaining terms, indicated as $O((q/k)^0)$, is not protected and will be modified in a less and less controllable way at increasing $k$ vaules and decreasing redshifts.

When the {\em hard} scale PS is evaluated at different times, $\tau_\beta\neq\tau_\gamma$, the BS in the squeezed limit goes as $P^{\rm lin}(q)/q$, and in real space ($f=0$) the contribution to the BS is a dipole, as it is proportional to $\mu/b_\alpha$. Physically, this contribution can be interpreted as the effect of the different large scale displacements, $\tilde\bd(\bq,\tau_\beta)$ and  $\tilde\bd(\bq,\tau_\gamma)$ experienced by the two short-scale fields $\delta^{(S)}_\beta(\bk,\tau_\beta)$, and $\delta^{(S)}_\gamma(\bk,\tau_\gamma)$ at the two different times $\tau_\beta$ and $\tau_\gamma$. This effect grows with the coherence length of the displacement, which explains the $1/q$ behavior, and moreover it depends on the orientation between the large scale and the short scale modes, which explains the dipole behavior.

The coefficient in front of the PS's goes (again, in real space) as $-\mu^2/b_\alpha \times O((q/k)^0)$. The $\mu^2$ behavior can be understood as follows. As we have already recalled, see Eq.~\re{etimes}, a perfectly uniform displacement field cannot give any contribution to the equal times BS. Therefore, the effect can depend only on the gradient of the large scale displacement/velocity field. More precisely, the $i-th$ spatial component of the displacement field can affect the clustering on short scales along the $j-th$ direction only via its $\partial_j d^i(\bx)$ component, leading to a contribution to the configuration space three point function proportional to
\beqra
&&  \langle \delta_\alpha(-{\bf R}) \delta_\alpha\left(\frac{\bf r}{2}\right) \delta_\alpha\left(-\frac{\bf r}{2}\right)\rangle \nonumber\\
&& \qquad\;\;\,=\langle \bar\delta_\alpha(-{\bf R}+\bd(-{\bf R})) \bar\delta_\alpha\left(\frac{\bf r}{2} +\bd\left(\frac{\bf r}{2} \right) \right) \bar\delta_\alpha\left(-\frac{\bf r}{2} +\bd\left(-\frac{\bf r}{2} \right) \right)  \rangle\nonumber\\
&& \qquad\;\;\,=  r^j\frac{\partial \xi_\alpha(r)}{\partial r^i}\,  \langle \delta_\alpha(-{\bf R})\left.\frac{ \partial d^i(\br)}{\partial r^j}\right|_{\br=0}\rangle +\cdots\,,\nonumber\\
&& \qquad\;\;\,= \frac{2}{3 {\cal H}^2}\frac{r^ir^j}{r^2}\frac{\partial \xi_\alpha(r)}{\partial \ln r}\,  \langle \delta_\alpha(-{\bf R})\left.\frac{ \partial^2 \Phi(\br)}{\partial r^i\partial r^j}\right|_{\br=0}\rangle +\cdots\,
\eeqra
where $\xi_\alpha(r)$ is the correlation function and $  \Phi(\br)$ the gravitational potential, which has been related to the displacement $\bd$ by means of linear PT.
 In Fourier space (see Eq.~\re{lt}), $\partial_j d^i(\br)$ gives $-q^iq^j/q^2\times \delta_m(\bq)$ which, contracted to $k^i\partial P_\alpha(k)/\partial k^j=(k^ik^j)/k^2\, d P_\alpha(k)/ d \ln k$ , gives the $-\mu^2$ dependence.

 The reason of the $1/b_\alpha$ factor comes about because we want to trade the velocity field (which is responsible for the CR protected term) with the directly observable density field for the $\alpha$ tracer.

\section*{References}
\bibliographystyle{JHEP}
\bibliography{CR_LSS.bbl}

\providecommand{\href}[2]{#2}\begingroup\raggedright\begin{thebibliography}{10}

\bibitem{Peloso:2013zw}
M.~Peloso and M.~Pietroni, {\it {Galilean invariance and the consistency
  relation for the nonlinear squeezed bispectrum of large scale structure}},
  {\em JCAP} {\bf 1305} (2013) 031 [\href{http://arXiv.org/abs/1302.0223}{{\tt
  1302.0223}}].

\bibitem{Kehagias:2013yd}
A.~Kehagias and A.~Riotto, {\it {Symmetries and Consistency Relations in the
  Large Scale Structure of the Universe}},  {\em Nucl.Phys.} {\bf B873} (2013)
  514--529 [\href{http://arXiv.org/abs/1302.0130}{{\tt 1302.0130}}].

\bibitem{Marinucci:2019wdb}
M.~Marinucci, T.~Nishimichi and M.~Pietroni, {\it {Measuring Bias via the
  Consistency Relations of the Large Scale Structure}},  {\em Phys. Rev.} {\bf
  D100} (2019), no.~12 123537 [\href{http://arXiv.org/abs/1907.09866}{{\tt
  1907.09866}}].

\bibitem{Peloso:2013spa}
M.~Peloso and M.~Pietroni, {\it {Ward identities and consistency relations for
  the large scale structure with multiple species}},  {\em JCAP} {\bf 1404}
  (2014) 011 [\href{http://arXiv.org/abs/1310.7915}{{\tt 1310.7915}}].

\bibitem{Creminelli:2013poa}
P.~Creminelli, J.~Gleyzes, M.~Simonovi{\'c} and F.~Vernizzi, {\it {Single-Field
  Consistency Relations of Large Scale Structure. Part II: Resummation and
  Redshift Space}},  {\em JCAP} {\bf 1402} (2014) 051
  [\href{http://arXiv.org/abs/1311.0290}{{\tt 1311.0290}}].

\bibitem{Valageas:2016hhr}
P.~Valageas, A.~Taruya and T.~Nishimichi, {\it {Consistency relations for large
  scale structures with primordial non-Gaussianities}},  {\em Phys. Rev. D}
  {\bf 95} (2017), no.~2 023504 [\href{http://arXiv.org/abs/1610.00993}{{\tt
  1610.00993}}].

\bibitem{Esposito:2019jkb}
A.~Esposito, L.~Hui and R.~Scoccimarro, {\it {Nonperturbative test of
  consistency relations and their violation}},  {\em Phys. Rev. D} {\bf 100}
  (2019), no.~4 043536 [\href{http://arXiv.org/abs/1905.11423}{{\tt
  1905.11423}}].

\bibitem{Creminelli:2013nua}
P.~Creminelli, J.~Gleyzes, L.~Hui, M.~Simonovi{\'c} and F.~Vernizzi, {\it
  {Single-Field Consistency Relations of Large Scale Structure. Part III: Test
  of the Equivalence Principle}},  {\em JCAP} {\bf 1406} (2014) 009
  [\href{http://arXiv.org/abs/1312.6074}{{\tt 1312.6074}}].

\bibitem{Eis05}
{\bf SDSS} Collaboration, D.~J. Eisenstein {\em et.~al.}, {\it {Detection of
  the Baryon Acoustic Peak in the Large-Scale Correlation Function of SDSS
  Luminous Red Galaxies}},  {\em Astrophys. J.} {\bf 633} (2005) 560--574
  [\href{http://arXiv.org/abs/astro-ph/0501171}{{\tt astro-ph/0501171}}].

\bibitem{Alam:2016hwk}
{\bf BOSS} Collaboration, S.~Alam {\em et.~al.}, {\it {The clustering of
  galaxies in the completed SDSS-III Baryon Oscillation Spectroscopic Survey:
  cosmological analysis of the DR12 galaxy sample}},  {\em Mon. Not. Roy.
  Astron. Soc.} {\bf 470} (2017), no.~3 2617--2652
  [\href{http://arXiv.org/abs/1607.03155}{{\tt 1607.03155}}].

\bibitem{Slepian_2017}
Z.~Slepian {\em et.~al.}, {\it {Detection of baryon acoustic oscillation
  features in the large-scale three-point correlation function of SDSS BOSS
  DR12 CMASS galaxies}},  {\em Mon. Not. Roy. Astron. Soc.} {\bf 469} (2017),
  no.~2 1738--1751 [\href{http://arXiv.org/abs/1607.06097}{{\tt 1607.06097}}].

\bibitem{Tyson:2002nh}
J.~Tyson, D.~Wittman, J.~Hennawi and D.~Spergel, {\it {LSST: A Complementary
  probe of dark energy}},  {\em Nucl. Phys. B Proc. Suppl.} {\bf 124} (2003)
  21--29 [\href{http://arXiv.org/abs/astro-ph/0209632}{{\tt
  astro-ph/0209632}}].

\bibitem{Amendola:2016saw}
L.~Amendola {\em et.~al.}, {\it {Cosmology and fundamental physics with the
  Euclid satellite}},  {\em Living Rev. Rel.} {\bf 21} (2018), no.~1 2
  [\href{http://arXiv.org/abs/1606.00180}{{\tt 1606.00180}}].

\bibitem{Dore:2019pld}
O.~Dor{\'e} {\em et.~al.}, {\it {WFIRST: The Essential Cosmology Space
  Observatory for the Coming Decade}},
  \href{http://arXiv.org/abs/1904.01174}{{\tt 1904.01174}}.

\bibitem{Taruya:2010mx}
A.~Taruya, T.~Nishimichi and S.~Saito, {\it {Baryon Acoustic Oscillations in
  2D: Modeling Redshift- space Power Spectrum from Perturbation Theory}},  {\em
  Phys. Rev.} {\bf D82} (2010) 063522
  [\href{http://arXiv.org/abs/1006.0699}{{\tt 1006.0699}}].

\bibitem{Kaiser:1987qv}
N.~Kaiser, {\it {Clustering in real space and in redshift space}},  {\em
  Mon.Not.Roy.Astron.Soc.} {\bf 227} (1987) 1--27.

\bibitem{Springel:2005mi}
V.~Springel, {\it {The Cosmological simulation code GADGET-2}},  {\em
  Mon.Not.Roy.Astron.Soc.} {\bf 364} (2005) 1105--1134
  [\href{http://arXiv.org/abs/astro-ph/0505010}{{\tt astro-ph/0505010}}].

\bibitem{Ade:2015xua}
{\bf Planck} Collaboration, P.~A.~R. Ade {\em et.~al.}, {\it {Planck 2015
  results. XIII. Cosmological parameters}},  {\em Astron. Astrophys.} {\bf 594}
  (2016) A13 [\href{http://arXiv.org/abs/1502.01589}{{\tt 1502.01589}}].

\bibitem{Scoccimarro:1997gr}
R.~Scoccimarro, {\it {Transients from initial conditions: a perturbative
  analysis}},  {\em Mon. Not. Roy. Astron. Soc.} {\bf 299} (1998) 1097
  [\href{http://arXiv.org/abs/astro-ph/9711187}{{\tt astro-ph/9711187}}].

\bibitem{Crocce:2006ve}
M.~Crocce, S.~Pueblas and R.~Scoccimarro, {\it {Transients from Initial
  Conditions in Cosmological Simulations}},  {\em Mon.Not.Roy.Astron.Soc.} {\bf
  373} (2006) 369--381 [\href{http://arXiv.org/abs/astro-ph/0606505}{{\tt
  astro-ph/0606505}}].

\bibitem{Nishimichi:2008ry}
T.~Nishimichi {\em et.~al.}, {\it {Modeling Nonlinear Evolution of Baryon
  Acoustic Oscillations: Convergence Regime of N-body Simulations and Analytic
  Models}},  {\em Publ. Astron. Soc. Jap.} {\bf 61} (2009) 321
  [\href{http://arXiv.org/abs/0810.0813}{{\tt 0810.0813}}].

\bibitem{Valageas:2010yw}
P.~Valageas and T.~Nishimichi, {\it {Combining perturbation theories with halo
  models}},  {\em Astron. Astrophys.} {\bf 527} (2011) A87
  [\href{http://arXiv.org/abs/1009.0597}{{\tt 1009.0597}}].

\bibitem{Nishimichi:2018etk}
T.~Nishimichi {\em et.~al.}, {\it {Dark Quest. I. Fast and Accurate Emulation
  of Halo Clustering Statistics and Its Application to Galaxy Clustering}},
  {\em Astrophys. J.} {\bf 884} (2019) 29
  [\href{http://arXiv.org/abs/1811.09504}{{\tt 1811.09504}}].

\bibitem{Behroozi:2011ju}
P.~S. Behroozi, R.~H. Wechsler and H.-Y. Wu, {\it {The Rockstar Phase-Space
  Temporal Halo Finder and the Velocity Offsets of Cluster Cores}},  {\em
  Astrophys. J.} {\bf 762} (2013) 109
  [\href{http://arXiv.org/abs/1110.4372}{{\tt 1110.4372}}].

\bibitem{hockney81}
R.~W. {Hockney} and J.~W. {Eastwood}, {\em {Computer Simulation Using
  Particles}}.
\newblock Taylor and Francis, 1981.

\bibitem{Jing:2004fq}
Y.~P. Jing, {\it {Correcting for the alias effect when measuring the power
  spectrum using FFT}},  {\em Astrophys. J.} {\bf 620} (2005) 559--563
  [\href{http://arXiv.org/abs/astro-ph/0409240}{{\tt astro-ph/0409240}}].

\bibitem{Sefusatti:2015aex}
E.~Sefusatti, M.~Crocce, R.~Scoccimarro and H.~Couchman, {\it {Accurate
  Estimators of Correlation Functions in Fourier Space}},  {\em Mon. Not. Roy.
  Astron. Soc.} {\bf 460} (2016), no.~4 3624--3636
  [\href{http://arXiv.org/abs/1512.07295}{{\tt 1512.07295}}].

\bibitem{ForemanMackey:2012ig}
D.~Foreman-Mackey, D.~W. Hogg, D.~Lang and J.~Goodman, {\it {emcee: The MCMC
  Hammer}},  {\em Publ. Astron. Soc. Pac.} {\bf 125} (2013) 306--312
  [\href{http://arXiv.org/abs/1202.3665}{{\tt 1202.3665}}].

\bibitem{Lewis:2019xzd}
A.~Lewis, {\it {GetDist: a Python package for analysing Monte Carlo samples}},
  \href{http://arXiv.org/abs/1910.13970}{{\tt 1910.13970}}.

\bibitem{Scoccimarro_1998}
R.~Scoccimarro, S.~Colombi, J.~N. Fry, J.~A. Frieman, E.~Hivon and A.~Melott,
  {\it Nonlinear evolution of the bispectrum of cosmological perturbations},
  {\em The Astrophysical Journal} {\bf 496} (Apr, 1998) 586--604.

\bibitem{Scoccimarro:1995if}
R.~Scoccimarro and J.~Frieman, {\it {Loop corrections in nonlinear cosmological
  perturbation theory}},  {\em Astrophys.J.Suppl.} {\bf 105} (1996) 37
  [\href{http://arXiv.org/abs/astro-ph/9509047}{{\tt astro-ph/9509047}}].

\end{thebibliography}\endgroup
\end{document}